\documentclass{article}

\usepackage{arxiv}

\usepackage[utf8]{inputenc} 
\usepackage[T1]{fontenc}    
\usepackage{hyperref}       
\usepackage{url}            
\usepackage{booktabs}       
\usepackage{amsfonts}       
\usepackage{nicefrac}       
\usepackage{microtype}      
\usepackage{graphicx}
\usepackage{xcolor}
\usepackage{mathtools}
\usepackage{amsmath}
\usepackage{amsthm}
\usepackage{siunitx}        
\usepackage[skip=0.33\baselineskip]{caption} 
\usepackage{multirow}
\graphicspath{{./Figures}}

\title{Estimating Causal Treatment Effects in Placebo-Controlled Randomized Clinical Trials When High Placebo Response is Anticipated}

\author{
  Yang Song \\
  Smith Center for Outcomes Research in Cardiology, Beth Israel Deaconess Medical Center\\
  Boston, MA, United States \\
  \texttt{ysong5@bidmc.harvard.edu} \\
  \And
  Yunzhe Qian \\
  Smith Center for Outcomes Research in Cardiology, Beth Israel Deaconess Medical Center\\
  Boston, MA, United States \\
  \And
  Chanmin Kim \\
  Department of Statistics, SungKyunKwan University\\
  Seoul, South Korea \\
  \And
  Gheorghe Doros \\
  Biostatistics Department, Boston University\\
  Boston, MA, United States \\
}

\begin{document}
\maketitle

\begin{abstract}
In placebo-controlled randomized clinical trials (RCTs), the placebo response significantly modifies treatment effects and diminishes the intention-to-treat (ITT) treatment effect, $\Delta_{ITT}$. This study presents a novel two-stage framework for estimating the standardized causal treatment effect, $\Delta_{STD}$, among the ITT population, under the assumption that their placebo responses are similar to the levels of self-administering medication at home. The first stage employs a real-world, pragmatic, single-blinded placebo lead-in to measure placebo responses to levels expected during routine at-home use. This is achieved by preserving the participants' expectations and controlling for trial-related factors that inflate the responses. In the second stage, a double-blinded randomized phase is used to estimate the conditional average treatment effect (CATE) as a function of placebo response levels and other important effect modifiers. To facilitate CATE estimation, the prognostic scores, defined as the expected placebo responses, are used for dimension reduction. The causal estimand $\Delta_{STD}$ is computed by integrating the CATE function over the distribution of the expected placebo response levels from stage one and other modifiers. We further derive theoretical values for $\Delta_{ITT}-\Delta_{STD}$ to quantify the underestimated treatment benefit due to high placebo responses. The validity and statistical performance of the proposed framework are evaluated through comprehensive simulations.
\end{abstract}


\section{Introduction}
\label{sec:Section1}
A high placebo response significantly reduces treatment effect and is a primary reason for trial failures. The placebo response serves as a strong effect modifier, leading to substantial heterogeneous treatment effects (HTE) across different response levels, causing treatment effects to diminish as response levels increase. When a high placebo response is evident in randomized controlled trials (RCTs), employing the intent-to-treat (ITT) treatment effect, which is derived from pooling various HTE estimates into a single value, complicates the regulatory review. More importantly, in everyday medicine use, it is unrealistic to expect trial participants to exhibit the same high levels of response as observed in the RCT. This discrepancy renders the ITT treatment effect less informative and not generalizable for understanding the treatment benefits of the same population if they were to use the same active treatment outside the study.

Researchers have proposed novel study designs and innovative statistical methods, or a hybrid of both approaches, aimed at estimating treatment effects by giving higher weights to subjects with lower placebo responses. 

Fava et al. introduced an innovative two-stage study design known as the sequential parallel comparison design (SPCD) \cite{Fava2003}. In the first stage, all ITT subjects are randomized to either active treatment or placebo, with a larger proportion assigned to the placebo arm. Placebo responders and non-responders are identified after this stage based on a pre-specified clinical threshold. In the second stage, placebo non-responders are re-randomized to receive either active treatment or placebo. This enables estimation of the treatment effect specifically within the non-responder subgroup. The overall treatment effect for the SPCD is derived by combining the first-stage estimate from the full ITT population with the second-stage estimate from the non-responders using a pre-specified weighted average approach. This design amplifies the treatment effect among placebo non-responders while attenuating the influence of placebo responders, thereby enhancing the sensitivity of the trial to detect a true treatment effect.

Gomeni et al. \cite{Gomeni2023} proposed a method for estimating treatment effects in RCTs with classical parallel design using a weighted mixed-effects model. In their approach, individual weights are defined as the inverse probability of being a placebo responder. This weighting scheme effectively down-weights participants who are more likely to be placebo responders, ensuring that responders have a reduced influence on the estimation of the overall treatment effect.

Building upon the strengths of both the SPCD and weighted regression methodologies, Rybin et al. \cite{Rybin2015} introduced an analytical framework that conceptualizes the two stages of the SPCD as separate parallel-design randomized controlled trials. Within each stage, the treatment effect is estimated via weighted least squares, where individual weights are defined as the estimated probability of being a placebo non-responder. The stage-specific estimates are then combined using pre-specified weights to derive an overall treatment effect, enhancing precision and robustness of the estimated treatment effect by prioritizing data from placebo non-responders.

Essentially, all these methods employ different group-level or individual-level weights to minimize the influence of responders on the estimation of the treatment effect. However, while weighting is a standard technique for causal inference, not all weighted averages are interpretable. Without a rigorous causal framework to guide the weighting strategy, it remains difficult to justify whether a given method accurately targets a meaningful and interpretable causal estimand.

This paper proposes a framework for estimating the causal treatment effect in randomized clinical studies where a high placebo response is anticipated. Section \ref{sec:Section1} outlines the challenge of high placebo response in RCTs, reviews existing weighting methods to mitigate its impact, and emphasizes the need for a formal causal framework to define interpretable treatment effects. Section \ref{sec:Section2} details the two-stage framework, which involves a real-world, pragmatic, single-blinded placebo lead-in phase to measure naturalistic placebo response levels and a double-blinded phase to estimate the CATE function. Section \ref{sec:Section3} formalizes the causal assumptions, provides a new look at the estimated placebo response, and rigorously defines the causal estimand $\Delta_{STD}$. Additionally, it gives a generic identification result for $\Delta_{STD}$ and presents a theoretical theorem that facilitates estimation of the underestimated treatment benefit in the ITT effect. Finally, Section \ref{sec:Section4} evaluates the performance of the proposed method through simulations under both low- and high-dimensional settings. This evaluation compares the absolute bias and mean squared error (MSE) for both $\Delta_{ITT}$ and $\Delta_{STD}$ across parametric, semi-parametric, and non-parametric estimation approaches.

\section{A Novel Two-Stage Framework}
\label{sec:Section2} 
\begin{figure}
    \centerline{\includegraphics[width=15.41cm, height=8cm]{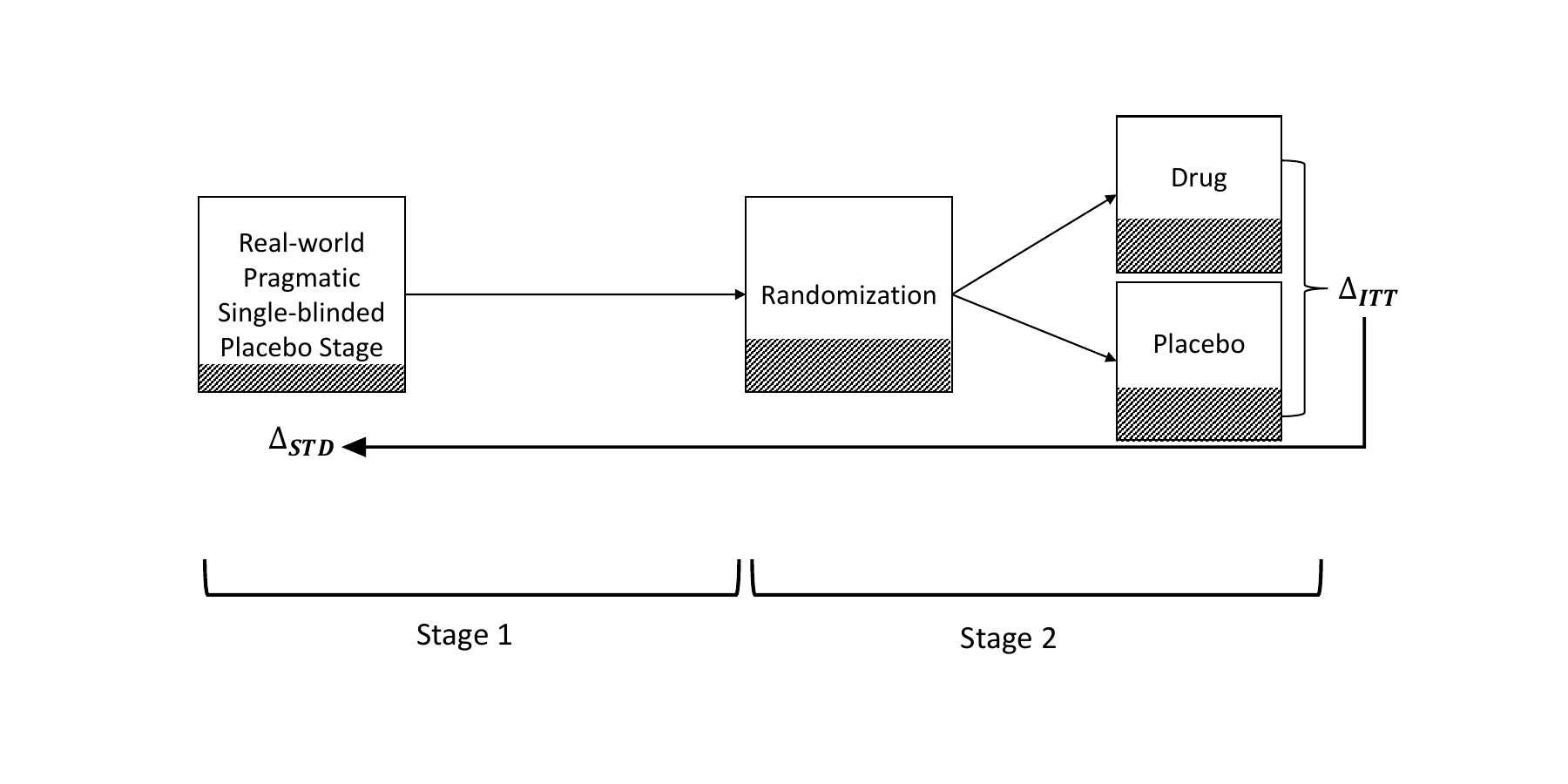}}
    \caption{\textbf{The Novel Two-stage Study Design.} $\Delta_{ITT}$ is attenuated due to the influence of high placebo responses (depicted as shaded regions) in both treatment arms. The proposed causal estimand, denoted $\Delta_{STD}$, represents the transported causal treatment effect for the same ITT population, under the more realistic response conditions observed in stage one.}
    \label{fig:fig1}
\end{figure}

As with other placebo-controlled RCTs, a central objective is to estimate the causal effect of active treatment compared to placebo within the ITT population, under the assumption that the level of placebo response observed in the trial is comparable to that expected during routine, self-administered use at home. To achieve this, we propose a novel two-stage design consisting of an initial real-world, pragmatic, single-blinded placebo lead-in phase (RW) followed by a classical double-blinded parallel comparison stage (DB). The study design is illustrated in Figure \ref{fig:fig1}. 

Throughout this article, the term "placebo response" refers to the improvement in symptom severity scores observed in patients who receive a placebo. This response is attributed to non-pharmacologic factors, such as patient expectations and the therapeutic setting. Placebo response here is a counterfactual concept that any individual included in the study has certain level of response if assigned to the deceptive placebo, but is only observed if the patient received a placebo, while remains latent in subjects who received the active treatment. 

In the first stage, all participants in the ITT population (or an even broader cohort) enter a single-blind phase where they receive a deceptive placebo. This phase is designed pragmatically to mimic typical at-home medication use and environment. This phase maintains participant expectation of receiving an active treatment through the use of deceptive placebo. To ensure the measured placebo response reflects conditions that are as close as possible to real-world situations, we control for key amplifiers of the placebo response within the trial setting, including visit frequency, duration of patient-clinician interaction, and structured patient education \cite{Rutherford2013,Fava2003}. The change in clinical severity scores observed at the end of this stage thus constitutes a proxy for the subject's latent placebo response level. We argue that by employing a deceptive placebo while controlling for trial-induced contextual factors, this measured response more accurately represents the hypothetical placebo response level expected during at-home self-administration.

The second stage consists of a double-blinded, randomized study. Rather than targeting the conventional ITT effect ($\Delta_{ITT}$), which is attenuated by high placebo responses in both treatment arms, we aim to estimate the Conditional Average Treatment Effect (CATE) as a function of placebo response levels and other baseline modifiers. As illustrated in Figure \ref{fig:fig1}, $\Delta_{ITT}$ is diluted by these high placebo responses (shaded regions). We therefore propose a standardized causal estimand, $\Delta_{STD}$, which transports the ITT effect in the trial to a treatment effect among the same ITT population, but under a realistic setting of real-world use of the study medication. Formally, $\Delta_{STD}$ is derived by integrating the CATE function over the distribution of placebo responses and other modifiers from the first stage. If the first-stage placebo responses indeed approximate those under real-world, self-administered use, this standardization provides a more interpretable and generalizable measure of treatment efficacy.\newline

\noindent \textbf{Justification of the Real-world, Pragmatic, Single-blinded Placebo Lead-in Phase}\newline
The initial stage aims to estimate the individual-level counterfactual placebo response that would occur during real-world use of the active medication. This represents the symptom reduction attributable to the mind-body processes of receiving treatment, rather than the drug's direct pharmacological effect. As this response is unobservable, where physicians do not prescribe deceptive placebos in real-world clinical practice, we employ this first stage to directly characterize this latent variable. 

A critical methodological question, therefore, is which study design can best achieve this goal. A review of the literature reveals two candidate designs: an open-label placebo lead-in and a single-blind placebo lead-in.

One option is to use the pragmatic single-blinded placebo \cite{Kaptchuk2010, Charlesworth2017} as a lead-in phase, in which, participants are knowingly consuming an inert substance prior to the second-stage double-blinded randomization. This approach prioritizes ethical transparency and provides a direct measure of placebo susceptibility under the no deception conditions.

Alternatively, the single-blind placebo lead-in employs a traditional "washout" design where participants believe they are receiving active treatment but are administered a placebo \cite{Trivedi1994}. This design engages patient expectancy, a core placebo mechanism, and is often used to identify and exclude "placebo responders" to enhance the assay sensitivity of the subsequent trial phase \cite{Scott2021}.

However, while conceptually aligned with our aim, both candidate designs possess critical limitations for estimating a real-world counterfactual placebo responses. The open-label placebo design, while ethically robust, alters a key component of real-world treatment: patient expectation. By explicitly removing this belief through full transparency, this design likely underestimates the placebo response present in actual clinical practice, where expectation is a powerful driver \cite{Papakostas2009,Benedetti2003}. Conversely, the single-blind placebo lead-in phase, conducted within the intensified context of an explanatory trial, with frequent monitoring and heightened clinician attention, artificially inflates the placebo response beyond what occurs in daily practice \cite{Enck2011,Severus2012}. The response is amplified by the "super-charged" therapeutic environment of the trial itself.

Therefore, neither standard designs can be directly estimate a valid proxy for the placebo response in real-world medication use, requiring a new method to approximate this counterfactual.

To address these limitations, we propose a novel first-stage design that integrates a real-world, pragmatic trial approach with a single-blind placebo lead-in. The core rationale is to preserve the crucial element of patient belief while mitigating the artificial inflation caused by an intensive trial context. A pragmatic trial aims to evaluate effectiveness under routine clinical conditions \cite{Ford2016}. By mirroring everyday clinical practice in this first-stage design, we deliberately de-intensify the trial setting to capture a more valid placebo response in the real-world setting. 

As a template example, the detailed steps for this stage are as follows:
\begin{enumerate}
    \item \textbf{Recruitment and Screening}: Although the use of broad eligibility criteria in a pragmatic trial is intended to create a real-world representative population and enhance the generalizability of the treatment effect, resource limitations can force a reduction in enrollment scope to as narrow as the ITT population.

    \item \textbf{Informed Consent Process}: The consent form should be transparent about the two-stage design while maintaining the blind for the stage-one. It will state, for example: "You are invited to participate in a study to understand how people respond to a new treatment strategy for the XYZ condition. This study has two parts. In this first part, all participants will receive a starter dose of medication to establish a baseline. In the second part, you will be randomly assigned to continue with one of two treatment paths."
    
    \item \textbf{Clinical Prescription}: A clinician will prescribe the study medication in a regular outpatient visit. Using a plausible dummy drug name, the clinician can script: "Based on your symptoms, I am prescribing you [the dummy drug name], which is often helpful for conditions like yours. It may take a few weeks to see the full effect." And no extra therapeutic encouragement should be provided. 
    
    \item \textbf{Pragmatic Follow-up and Outcome Assessment}: Participants will be dispensed inert pills labeled with the dummy drug name. Outcome data will be collected using low-burden methods (e.g., patient diaries, simple pill counts). To maintain a pragmatic feel, there will be no protocol-mandated interim contacts. Participants are given a standard clinic phone number to call with questions or concerns, replicating the procedures in normal clinical practice.
    
    \item \textbf{End of Lead-in and Calculation of Placebo Response}: At the end of the lead-in period, the primary outcome (typically a severity score) will be assessed. The individual-level placebo response will then be calculated for each participant as the change from baseline in this score.
    
    \item \textbf{Transition to Stage 2}: Participants will be fully debriefed about the first stage and re-consented for Stage 2. The interviewing script will explain: "The first medication was a placebo, an inactive pill, which was necessary to understand your personal response to starting a new treatment. We now invite you to continue to the second stage, where you will be randomly assigned to receive either the real active medication or a matched placebo." Participants who re-consent will be formally randomized into the double-blinded phase.
\end{enumerate}

This pragmatic single-blinded placebo lead-in design directly targets our primary aim. It preserves the crucial element of belief (which the open-label placebo design eliminates) while mitigating the artificial inflation of the placebo response (the core weakness of the traditional trial-based single-blinded run-in). The resulting change in the primary outcome (e.g., the severity score) provides a more valid approximation of the latent, expectation-driven symptom reduction embedded within the real-world use of any active medication. This individual-level variable will subsequently be used as key information to precisely isolate the specific pharmacological effect of the drug from the non-specific placebo responses.

\section{Statistical Methods}
\label{sec:Section3}
This section formalizes the statistical methods to identify the $\Delta_{STD}$ using the novel two-stage framework. In Section \ref{sec:Section3.1}, we begin by presenting the causal DAG and stating the assumptions. Next, we reexamined the placebo response and defined the expected placebo response as a prognostic score, which serves as a key effect modifier to simplify the CATE function in Section \ref{sec:Section3.2}. We then define the causal estimand $\Delta_{STD}$ in Section \ref{sec:Section3.3} and provide its identification in Section \ref{sec:Section3.4}. To identify $\Delta_{STD}$, we demonstrated the identification process through both generic steps and a theoretical derivation under specific parametric assumptions. 

\subsection{Causal DAG and Assumptions}
\label{sec:Section3.1}
\begin{figure*}
    \centerline{\includegraphics[width=8cm, height=4cm]{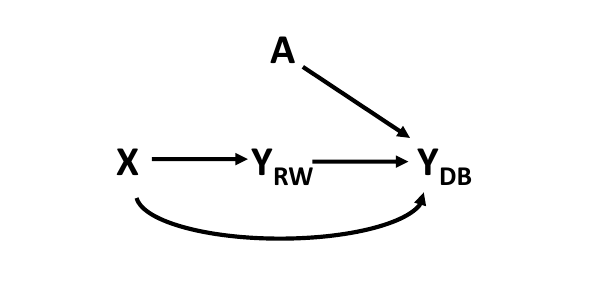}}
    \caption{Causal DAG of the Novel Two-Stage Framework}
    \label{fig:fig2}
\end{figure*}

Figure \ref{fig:fig2} uses a directed acyclic graph (DAG) to to depict the causal structure of the two-stage study. The node $X$ encompasses a comprehensive collection of patient-level, physician-level, and hospital-level variables at baseline, including the initial severity score $Y_0$.

In the first pragmatic stage, all ITT subjects are assigned a single-blind placebo. Their severity score at the end of this stage is denoted as $Y_{RW}$, and the change from baseline, $\Delta_{RW} = Y_{RW} - Y_0$, quantifies the level of placebo response exhibited during this stage.

At the start of the second stage, subjects are randomized in a double-blind manner to either active treatment ($A=1$) or placebo ($A=0$). The severity score $Y_{DB}$ is measured at the end of this stage. For participants assigned to the placebo group, the change in severity score $\Delta_{DB} = Y_{DB} - Y_{RW}$ reflects the placebo response observed during this stage. For the active treatment group, $\Delta_{DB}$ captures the combined effect of the placebo response and the pharmacological treatment. 

We assume that $X$ contains rich baseline information, such that it captures all predictors $p(X)$ for $\Delta_{RW}$ and $\Delta_{DB}$, as well as the effect modifiers $M:=m(X)$ for treatment effect $\Delta_{DB}^{(A=1)} - \Delta_{DB}^{(A=0)}$. 

Under the assumption that the first-stage placebo run-in does not affect the second-stage treatment effect, the ITT effect can be observed using the second stage data, as \(\Delta_{ITT} = E(\Delta_{DB}^{(A=1)} - \Delta_{DB}^{(A=0)})\). This effect will underestimate the true treatment benefit of the active treatment in a daily use setting, especially if a significant placebo response is observed in the improvement of the severity score \(\Delta_{DB}\) among those receiving the placebo. 

Let $R_{DB}$ be a generic representation of the levels of the placebo response exhibited during the double-blind phase. This placebo response could be defined as a function of absolute improvement, relative improvement, or a combination of both. The ITT treatment effect can then be reformulated as: \(\Delta_{ITT} = E\left\{E(\Delta_{DB}^{(A=1)} - \Delta_{DB}^{(A=0)} \mid R_{DB})\right\}\). Given that the placebo response significantly modifies the treatment effect, the effective treatment benefit among high placebo responders is notably reduced. Consequently, the diluted treatment effect among these high placebo responders is included in the estimation of the ITT treatment effect, which biases it toward the null.  

To address this limitation, we propose to estimate the standardized treatment effect \(\Delta_{STD}\) among the same ITT population, but instead of calculating the ITT treatment effect by averaging the conditional average treatment effect over $R_{DB}$, similar to the standardization process, we propose to estimate it over the placebo response manifested during the real-world pragmatic first-phase $R_{RW}$, expressed as \(\Delta_{STD} = E\left\{E(\Delta_{DB}^{(A=1)} - \Delta_{DB}^{(A=0)} \mid R_{RW})\right\}\).

To rigorously define the causal estimand \(\Delta_{STD}\), we outline the causal assumptions used. The generic outcome notation $Y$ is used for $\Delta_{DB}$. The foundational assumptions of consistency, exchangeability, and positivity are described as follows:

\begin{itemize}
\item \textbf{A1. Consistency}: This assumption bridges the counterfactual outcomes to the observed values. For each individual, the observed outcome \(Y_i\) is assumed to be the same in value as the counterfactual outcome \(Y_i^{(A_i=a)}\) if the individual was assigned to treatment \(A_i=a\). Expressed as: \(Y_i = A_i*Y_i^{(A_i=1)} + (1-A_i)*Y_i^{(A_i=0)}\).

\item \textbf{A2. Exchangeability}: For the treatment assignment variable \(A_i\), marginal exchangeability is satisfied due to randomization, expressed as: \(Y_i^{(A_i=a)} \perp A_i, \forall a\).  

\item \textbf{A3. Positivity}: This assumption ensures subjects randomized to each treatment arm have a probability $\in (0,1)$ for any measured covariates and functions of X: $h(X)$, and can be expressed as: \(0 < P(A_i=a \mid h(X)) < 1, \forall a, h(X).\)
\end{itemize}

Due to randomization, A1 to A3 are reasonable assumptions in a well-conducted RCT. To simplify the exposition, we assume perfect adherence to the assigned treatment \(A\), the absence of missing data in the baseline covariates, and no early dropout during the two study phases. 

Under assumptions A1 to A3, any parametric, semi-parametric, or non-parametric outcome regression models that use treatment assignment $A$ and baseline predictors $X$ to predict the outcome values $Y$, can be formulated as a regression function with two parts: $\Psi(X)$ - average outcomes given treatment $A=0$ and baseline covariates $X$, and $\Pi(X)$ - conditional average treatment effect (CATE) given baseline covariates $X$.
\begin{flalign}
E(Y \mid A,X) & = E\left[A*Y^{(A=1)} + (1-A)*Y^{(A=0)} \bigm| A,X\right] & \llap{(by A1)} \nonumber\\
              & = E\left[Y^{(A=0)} + A*(Y^{(A=1)}-Y^{(A=0)}) \bigm| A,X\right] \nonumber\\ 
              & = E\left[Y^{(A=0)} \bigm| A,X\right] + A*E\left[Y^{(A=1)}-Y^{(A=0)} \bigm| A,X\right] \nonumber \\
              & = E\left[Y^{(A=0)} \bigm| A=0,X\right] + A*E\left[Y^{(A=1)}-Y^{(A=0)} \bigm| X\right] & \llap{(by A2 and A3)} \nonumber \\
              & = E\left[Y \bigm| A=0,X\right] + A*E\left[Y^{(A=1)}-Y^{(A=0)} \bigm| X\right] & \llap{(by A1)} \nonumber \\
              & = \Psi(X) + A*\Pi(X) & \label{eq:eq1}
\end{flalign}

Estimating the functions $\Psi(X)$ or $\Pi(X)$ from data without parametric assumptions presents significant challenges. Among these, $\Pi(X)$ appears more manageable to learn since it is primarily influenced by effect modifiers.

Previous work has focused on estimating the CATE function $\Pi(X)$ using finite-dimensional parameters, while treating $\Psi(X)$ as an infinite-dimensional nuisance function. Robinson introduced a double residualization approach that identifies the treatment effect by nonparametrically partialling out the influence of covariates from both the outcome $Y$ and the treatment $A$, leaving a residualized relationship that characterizes $\Pi(X)$ \cite{Robinson1988}. Ai and Chen generalized this insight into a flexible Sieve Minimum Distance framework, formalizing estimation via series approximations and generalized method of moments \cite{Ai2003}. Building on these foundations, Chernozhukov et al. (2018) further extended the approach to accommodate modern machine learning methods, rather than classical kernels, for estimating high-dimensional nuisance functions. By incorporating Neyman-orthogonal moment conditions in the residualization step and using cross-fitting to avoid overfitting, their double/debiased machine learning framework provides robust and theoretically valid inference \cite{Chernozhukov2018}.

Despite these advances, fundamental limitations in consistency, as highlighted by Stone et al. \cite{Stone1982}, pose serious obstacles in finite samples. Even when the CATE function is smooth with p-bounded derivatives, consistent estimation becomes unattainable when the dimension $d = \text{dim}(X)$ exceeds $\log(N)$. RCTs, which often have limited sample sizes, face particular difficulty in estimating $\Pi(X)$ accurately. For instance, an RCT with 10,000 participants can only accommodate at most four effect modifiers without further assumptions.

Moreover, the standard CATE function $\Pi(X)$ does not fully align with our substantive objectives. Rather than estimating a potentially high-dimensional $\Pi(X)$, we propose a more structured and interpretable CATE of the form $\Pi(R, M)$, where $R$ denotes the placebo response and $M$ represents a set of key effect modifiers, even after controlling for the strong effect modifiers $R$. 

The estimation of the standardized treatment effect, which leverages the first-stage placebo response and the CATE function $\Pi(R, M)$ estimated from the second-stage data, relies on the following key assumption:
\begin{itemize}
\item \textbf{A4: Stage-Invariance of the CATE Function}: The CATE function $\Pi(R,M)$ is invariant to the trial stage. Specifically, the CATE function estimated in the second stage is applicable to the first-stage context and is not modified by the single-blinded placebo assignment. This is formalized by two sub-assumptions: 
    \begin{itemize}
        \item \textbf{Transportability of CATE}: $\Pi(R_{DB}=r, M)=\Pi(R_{RW}=r, M)$. This is the central assumption that allows us to use the second-stage CATE to estimate the standardized effect $\Delta_{STD}$ for the patients  had they exhibited the first-stage placebo response levels $R_{RW}$.
    
        \item \textbf{No Carry-Over Effect on CATE}: $\Pi(R_{DB}=r, M) = E[Y^{(A=1)}-Y^{(A=0)} \mid R_{DB}=r,M] = E[Y^{(A=1)}-Y^{(A=0)} \mid R_{RW}, R_{DB}=r,M]$. This assumption ensures that the first-stage, single-blind placebo assignment does not alter the CATE function in the second stage. This is necessary to obtain an unbiased estimate of the ITT effect ($\Delta_{ITT}$) from the second-stage data.
    \end{itemize} 
\end{itemize}

As a remark, a violation of the no carry-over effect assumption would most plausibly attenuate the second-stage placebo response ($R_{DB}$) due to the first-stage placebo exposure. In our design, this would lead to an overestimation of the ITT effect $\Delta_{ITT}$. The reason is that a diminished response will inflate the observed treatment-control difference. Crucially, this violation does not affect the estimation of the standardized effect ($\Delta_{STD}$), provided the transportability of the CATE assumption holds. Consequently, a violation of the no carry-over assumption would likely to increase the probability of demonstrating treatment efficacy using the second stage data to estimate $\Delta_{ITT}$ in our proposed design, while the estimate for $\Delta_{STD}$ would remain valid and unbiased under the transportability of CATE assumption.

Following the specification of causal assumptions, the subsequent task is the operationalization of the variable $R$ to ensure the CATE function $\Pi(R, M)$ is both estimable and interpretable. A well-defined placebo response variable $R$ serves a dual purpose: it reduces dimensionality while preserving clinical interpretability. This formulation is essential for the practical application of $\Pi(R, M)$ to transport treatment effects and estimate the standardized effect $\Delta_{STD}$.


\subsection{A New Look at the Placebo Response} 
\label{sec:Section3.2}
The placebo response, following the conceptualization of Rutherford et al.\cite{Rutherford2013}, refers to the change in symptoms observed among patients assigned to the placebo arm during a clinical trial. It constitutes a latent individual-level characteristic that manifests only under placebo assignment and becomes observable after a period of follow-up. The magnitude of this response is typically quantified through absolute or relative changes in clinical outcomes among placebo-assigned participants.

Let \(Y=\Delta_{DB}\) denote the absolute change in outcome during the double-blinded phase. Under the potential outcomes framework, the placebo response is formally defined as \(Y^{(A=0)}\), representing the change that would occur if a patient were assigned to placebo. For individuals in the placebo group, this value is observed directly by the consistency assumption. For those receiving active treatment, however, \(Y^{(A=0)}\) remains latent and must be estimated. Previous research indicates that baseline characteristics, such as milder symptoms, shorter disease duration, and more frequent patient--provider interactions, are associated with greater placebo responses \cite{Howe2019,Wong2022}.

The expected placebo response can naturally be defined using the prognostic score, denoted as \(\Psi(X) \coloneqq E(Y^{(A=0)}|X)\). This function serves as a sufficient statistic for the outcome \(Y^{(A=0)}\), such that \(Y^{(A=0)} \perp X \mid \Psi(X)\). A notable advantage of employing the prognostic score as an estimator for the placebo response is its direct correspondence to the first term in the equation (\ref{eq:eq1}). Additionally, since the prognostic score operates as a sufficient statistic for \(Y^{(A=0)}\), it effectively serves as a dimension reduction tool for \(Y^{(A=0)}\). This concept parallels the role of the propensity score, \(p(A \mid X)\), which functions as a sufficient statistic for treatment assignment \(A\), satisfying the condition \(A \perp X \mid p(A \mid X)\) (treatment assignment is independent of covariates given the propensity score).

As discussed by Hansen \cite{Hansen2008}, a sufficient statistic for \(Y^{(A=0)}\) is not necessarily sufficient for \(Y^{(A=1)}\). When effect modifiers \(M \subset X\) are present, the expression \(E(Y^{(A=1)} - Y^{(A=0)} \mid \Psi(X))\) is a function of \(M\), rather than a constant. Let \((\Psi(X), M)\) be jointly sufficient for \(Y^{(A=1)}\). Instead of estimating the high-dimensional CATE function \(\Pi(X)\), we can work with the lower-dimensional CATE function \(\Pi(\Psi(X), M)\). This reduction in dimensionality is supported by the fact that the expected placebo response $\Psi(X)$ is a strong effect modifier, after conditioning on it, the influence of many other weak effect modifiers becomes negligible. 

To estimate the placebo response for participants assigned to the active treatment arm, we need to develop a prognostic score model \(\hat{\Psi}(X)\) using data from the placebo arm. This model will then be applied to estimate the prognostic score for each individual in the active treatment group. The individual-level additive assumption posits that the overall improvement is the sum of the prognostic score and the effect of the active treatment. Consequently, for an individual who received treatment $(A=1)$, the individual treatment effect can be estimated as the difference between their observed outcome $Y$ and their estimated prognostic score: $Y^{(A=1)} - Y^{(A=0)} = Y - \hat{\Psi}(X)$.

Furthermore, we can also define the prognostic score for the first-stage among all ITT subjects in a similar manner as \(\Phi(X) \coloneqq E(\Delta_{RW} \mid X)\), such that \( \Delta_{RW} \perp X \mid \Phi(X) \). It is important to recognize that the true prognostic scores \(\Phi(X)\) at the pragmatic first-stage and \(\Psi(X)\) at the double-blinded phase may not share the same functional forms, meaning a given predictor may influence these two scores differently, both in magnitude and even in direction. Without considering the early dropouts, each ITT subject has an estimated placebo response $\hat{\Phi}(X)$ at the pragmatic phase as well as that in the double-blinded phase $\hat{\Psi}(X)$.\newline 

\noindent\textbf{How to Identify Other Effect Modifiers $M$?}\newline 
While the primary goal of this study is not the empirical identification of effect modifiers, advancing the proposed methodology requires a principled approach for specifying variables $M$ that modify the treatment effect conditional on the strong modifier $R$. For this goal, we outline a two-stage procedure. First, we leverage a priori field knowledge to prioritize a set of candidate effect modifiers, acknowledging that a dominant modifier like predicted placebo response will account for substantial heterogeneity. Subsequently, to robustly identify which of these candidates are active conditional effect modifiers, one would ideally employ machine learning tools for estimating the CATE, such as causal forests \cite{Athey2019} and meta-learners \cite{Knzel2019}. However, given the limited sample size of a single trial, such machine learning methods are likely underpowered. To address this power limitation, a more powerful alternative involves pooling data from a series of similar trials to robustly identify additional sources of effect heterogeneity. 

For the purpose of developing our method, we will proceed under the assumption that the set $M$ can be identified through such a combination of field knowledge and machine learning methods on the expanded data, allowing us to focus on the subsequent method development.

\subsection{Causal Estimand $\Delta_{STD}$}
\label{sec:Section3.3}
After estimating the placebo response for all participants in both stages, we leverage data from the double-blinded phase to learn the rules governing conditional average treatment effects. These rules are captured by the CATE function, which takes the placebo response level $R$ and other key effect modifiers $M$ as parameters. 

This function will later be utilized in the first-stage data to estimate the target causal estimand $\Delta_{STD}$. This estimand reflects the treatment effect of the active treatment compared to placebo among the ITT population, under the assumption that every individual demonstrates the same level of placebo response during the first phase. By deliberately creating an environment that simulates regular medication use at home during pragmatic single-blinded placebo phase, $\Delta_{STD}$ becomes a more informative causal estimand. It quantifies the active drug's treatment benefit (compared to the hypothetical prescription of a placebo) in a home setting, in contrast to the ITT treatment effect, which measures the same treatment effect but within the context of a randomized clinical trial.

To formally illustrate these concepts, we define the generic CATE function as: $\Pi(.) = E(Y^{(A=1)} - Y^{(A=0)} \mid .)$. For example, the CATE function $\Pi(X)$ in precision medicine is employed to customize personalized treatment benefits based on the characteristics of each individual, represented by $X$. It is important to note that while CATE is used to estimate the individual treatment effect, it still represents an average effect across a more targeted samples, characterized by $X$. In a one-dimensional space defined by the placebo response level $R$, the CATE function $\Pi(R=r) = E(Y^{(A=1)} - Y^{(A=0)} \mid R=r)$ reflects the average treatment effect for individuals exhibiting a placebo response level of $R=r$. When additional effect modifiers $M$ exist, $\Pi(R=r)$ varies across different values of $M$ and is no longer sufficient for characterizing $Y^{(A=1)}$. Consequently, a more detailed CATE function $\Pi(R,M)$ is required to accurately quantify the treatment effect.

Similar to the analyses using the propensity score stratification or matching approaches, the CATE function $\Pi(p(X)=p)=E(Y^{(A=1)} - Y^{(A=0)} \mid p(X)=p)$ is estimated as an intermediate step implicitly, prior to averaging across target populations for marginal treatment effect, which computes the average treatment effect based on the propensity score $p$. However, as noted by MaCurdy et al. \cite{MaCurdy2011}, there is often little interest in the quantity $\Pi(p(X)=p)$. Given that $X$ is multi-dimensional, individuals with the same propensity score $p(X)$ may exhibit considerable heterogeneity in their values of $X$, leading to diverse treatment effects.

To estimate $\Delta_{STD}$, it is essential to learn the CATE function in the dimensions spanned by $R$ and $M$, and subsequently apply it to the data collected during pragmatic first phase. Assuming that we have accurately identified the effect modifiers $M$, and accurately estimated the placebo response $R$ for both stages through the prognostic scores $R=\Phi(X)$ and $R=\Psi(X)$, then we can learn the CATE function $\Pi(R=\Psi(X),M)$ from the double-blind stage data using parametric, semi-parametric, or non-parametric methods. 

For the double-blinded phase, by averaging the CATE function sequentially across $M$ and then $R$, we can estimate the conditional treatment effect as a function of placebo response at the double-blinded phase $\Pi_{DB}(\Psi(X)=r)$, as well as the ITT treatment effect $\Delta_{ITT}$, as demonstrated in the equations below.
\begin{flalign}
\label{eq:eq2}
& \Pi_{DB}(\Psi(X)=r) = E\left(Y^{(A=1)} - Y^{(A=0)} \mid \Psi(X)=r \right) = \int\Pi(\Psi(X)=r,M=m)f_{M|\Psi(X)=r}(M=m)dm \\
\label{eq:eq3}
& \Delta_{ITT} = E\left[ E\left(Y^{(A=1)}- Y^{(A=0)} \mid \Psi(X)=r \right) \right] = \int\Pi_{DB}(\Psi(X)=r)f(\Psi(X)=r)dr
\end{flalign}

Under the assumption of CATE transportability, the function $\Pi(R, M)$ governs both phases of the design. This allows us to derive a conditional average treatment effect based on the first-stage placebo response, denoted $\Pi_{RW}(\Phi(X)=r)$. The standardized average treatment effect, $\Delta_{STD}$, is then obtained by averaging $\Pi_{RW}(\Phi(X)=r)$ over the distribution of the response levels $\Phi(X)$.
\begin{flalign}
\label{eq:eq4}
& \Pi_{RW}(\Phi(X)=r) = E\left(Y^{(A=1)} - Y^{(A=0)} \mid \Phi(X)=r \right) = \int\Pi(\Phi(X)=r,M=m)f_{M|\Phi(X)=r}(M=m)dm \\
\label{eq:eq5}
& \Delta_{STD} =  E\left[ E\left(Y^{(A=1)}- Y^{(A=0)} \mid \Phi(X)=r \right) \right] = \int\Pi_{RW}(\Phi(X)=r)f(\Phi(X)=r)dr 
\end{flalign}

As a technical note, in the absence of other effect modifiers $M$, the functions in equations \ref{eq:eq2} and \ref{eq:eq4} are equivalent: $\Pi_{DB}(\Psi(X)=r) = \Pi_{RW}(\Phi(X)=r)$, for any $r$. However, when effect modifiers $M$ is present, these functions generally differ. This discrepancy occurs because, although we assume $\Pi(\Psi(X)=r,M) = \Pi(\Phi(X)=r,M)$, the conditional distributions $M\mid\Psi(X)=r$ and $M\mid\Phi(X)=r$ may differ between phases. This is intuitive, as the relationship between patient characteristics and the placebo response can change from the pragmatic single-blinded to the double-blinded setting. This crucial distinction complicates the transportation of the average treatment effect $\Delta_{ITT}$ to $\Delta_{STD}$, by directly estimating the one-dimensional CATE function in equation \ref{eq:eq2} and then averaging over the distribution of $f(\Phi(X)=r)$.


\subsection{Identification of $\Delta_{STD}$} 
\label{sec:Section3.4}
This subsection outlines the general steps for identifying the causal estimand $\Delta_{STD}$, a process applicable to all parametric, semi-parametric, and non-parametric estimation methods. We then present a theorem for estimating $\Delta_{STD}$ from $\Delta_{ITT}$ under parametric assumptions for the CATE function $\Pi(R,M)$. This theorem provides the theoretical foundation for deriving $\Delta_{STD}$ and the two CATE functions, $\Pi_{DB}(\Psi(X)=r)$ and $\Pi_{RW}(\Phi(X)=r)$ in Equations \ref{eq:eq2} and \ref{eq:eq4}, in our simulation studies.

\subsubsection{Generic Steps to Estimate $\Delta_{STD}$}
To estimate the standardized treatment effect, $\Delta_{STD}$, we discretize the predicted placebo response, $\hat{\Phi}(X)$, from the pragmatic first phase into $k$ distinct strata ($r_1, r_2, ..., r_k$). Let $n_i$ be the number of individuals in stratum $r_i$, with total sample size $N = \sum_{i=1}^{k}n_i$. Using Equations \ref{eq:eq4} and \ref{eq:eq5}, we then approximate the probability $\hat{\Pi}_{RW}(\Phi(X)=r)$ and $\hat{\Delta}_{STD}$ as follows:

\begin{flalign}
\label{eq:eq6}
& \hat{\Pi}_{RW}(\Phi(X)=r_i) = \frac{1}{n_i} \sum_{X:\Phi(X)=r_i} \hat{\Pi}(\Phi(X)=r_i, M=m)  \\
\label{eq:eq7}
& \hat{\Delta}_{STD} = \sum_{r_i} \hat{\Pi}_{RW}(\Phi(X)=r_i) \frac{n_i}{N} = \frac{1}{N} \sum_{r_i} \sum_{X:\Phi(X)=r_i} \hat{\Pi}(\Phi(X)=r_i, M=m)
\end{flalign}

This approach rests on the following critical assumptions:

\begin{itemize}
    \item A predictive prognostic score $\hat{\Phi}(X)$ is well-established and calibrated from the initial-stage data.
    \item The CATE function $\hat{\Pi}(R, M)$ is accurately estimated from the double-blinded phase.
\end{itemize}

An accurate prognostic score $\hat{\Psi}(X)$ is crucial for estimating the CATE, as it is learned from the double-blinded phase. The estimation of $\Delta_{STD}$ therefore involves the following steps:
 
\begin{enumerate}
\item Develop the prognostic score $\hat{\Psi}(X)$: Using data from the placebo arm of the double-blinded phase, develop the prognostic score model $\hat{\Psi}(X)$ based on the baseline patient, physician, and hospital characteristics. Apply this fitted model to subjects in the active treatment arm.
\item Estimate the CATE function $\hat{\Pi}(R, M)$: Using data from the active treatment arm of the double-blinded phase, estimate the CATE function, $\hat{\Pi}(R=\hat{\Psi}(X), M)$.
\item Develop the prognostic score $\hat{\Phi}(X)$ for the pragmatic first phase: Using data from the real-world pragmatic single-blinded placebo phase, develop the prognostic score model $\hat{\Phi}(X)$ for all ITT subjects. 
\item Compute $\hat{\Delta}_{STD}$: Discretize the range of $\hat{\Phi}(X)$ and apply Equations \ref{eq:eq6} and \ref{eq:eq7} to the resulting categories to estimate $\hat{\Delta}_{STD}$.
\end{enumerate}

As a technical note, the range of the first-stage prognostic score, $\Phi(X)$, must be a subset of the range of the double-blinded score, $\Psi(X)$, to avoid extrapolation when calculating $\Delta_{STD}$. We expect this condition to hold because the placebo response in real-world settings is typically lower and less variable than in double-blinded trials. This leads to a more centralized distribution for $\Phi(X)$, naturally confining its range within that of $\Psi(X)$. In practice, if an estimated value $\hat{\Phi}(X)$ falls outside the observed range of $\Psi(X)$, it should be handled during discretization by truncating it to the nearest boundary category of the discretized $\Psi(X)$.

\subsubsection{Theoretical Derivation Under Parametric Assumptions}
The four steps outlined above for the estimation of $\Delta_{STD}$ are applicable across parametric, semi-parametric, and non-parametric approaches for estimating $\Phi(X)$, $\Psi(X)$, and $\Pi(R,M)$. However, under parametric assumptions regarding $\Pi(R,M)$, the estimation process for $\Delta_{STD}$ can be significantly simplified. In this section, we present a theoretical derivation of the standardized treatment effects $\Delta_{STD}$, utilizing equations \ref{eq:eq2} through \ref{eq:eq5}. By positing a linear parametric assumption in the CATE function, this derivation facilitates the quantification of the underestimated treatment benefit ($\Delta_{ITT} - \Delta_{STD}$) due to the excessive placebo responses in the double-blinded phase. Additional assumptions are described formally as follows:

\begin{itemize}
\item B1: Let the $p$-dimensional random vector of baseline covariates be represented as $X = (X_1, X_2, ..., X_P)^T$, which are assumed to be independent and identically distributed across each individual $i$. The baseline severity score $Y_0$, the predictors of prognostic scores at each stage, and the effect modifiers are captured in $X$. We denote the true prognostic scores as $\Phi(X)$ and $\Psi(X)$, are defined as functions of $X$ that model the expected change in the severity score attributable to the placebo responses during the pragmatic single-blinded placebo lead-in and double-blinded phases, respectively.

\item B2: Assume that the CATE function is linear in the basis expansion of the placebo response $R$ and other effect modifiers $M$, such that $\Pi(R,M) = \pi^T B_R + \gamma^T M$. Here, $B_R$ is a basis function for $R$, and $M$ is a vector of effect modifiers. This formulation explicitly excludes higher-order interaction terms between $R$ and $M$. In Appendix \ref{sec:sectionD}, we present the complex CATE function when $R$ and $M$ interact, with both theoretical derivations and simulation studies to demonstrate the validity of the generic identification.
\end{itemize}

As a technical note, the basis $B_R$ can be constructed to be flexible and of higher order. A simple polynomial basis of order three is $B_R = (1, R, R^2, R^3)^T$, yielding $\pi^T B_R = \pi_0 + \pi_1 R + \pi_2 R^2 + \pi_3 R^3$. Alternatively, an orthogonal polynomial basis (e.g., Legendre polynomials up to order three) can be specified as: $B_R = (1, R, \frac{1}{2}(3R^2 - 1), \frac{1}{2}(5R^3 - 3R))^T$, leads to the expansion $\pi^T B_R = \pi_0 + \pi_1 R + \pi_2 \left[\frac{1}{2}(3R^2 - 1)\right] + \pi_3 \left[\frac{1}{2}(5R^3 - 3R)\right]$. The use of orthogonal polynomials is recommended to mitigate multicollinearity within the basis, which enhances the stability and accuracy of the parameter estimates $\pi$. This is particularly important when employing higher-order basis, where estimates can be highly sensitive to small data entry errors if the basis functions are strongly correlated.\newline

\noindent\textbf{Theorem (Quantification of the Underestimated Treatment Benefit)}\newline 
Given the additional assumptions B1 and B2, the underestimated treatment benefit due to excessive placebo response in the double-blinded phase of using $\Delta_{ITT}$ as study treatment effect is given by:
\begin{flalign}
\Delta_{ITT}-\Delta_{STD}= \pi^T[E_{\Psi}(B_R)-E_{\Phi}(B_R)] \nonumber
\end{flalign}
where $E_{\Phi}(B_R)$ denotes the expectation of $B_R$ with respect to the distribution of the prognostic score $\Phi(X)$ from the pragmatic first phase, and $E_{\Psi}(B_R)$ denotes the expectation with respect to the distribution of $\Psi(X)$ from the double-blinded phase.

\begin{proof}
Let $M=(M_1, M_2, ..., M_m)^T$ be the vector of effect modifiers. For the double-blinded stage, the CATE as a function of placebo response level can be derived as:
\begin{flalign}
\Pi_{DB}(\Psi(X)=r) & = \int\Pi(\Psi(X)=r,M=m)f_{M|\Psi(X)=r}(M=m)dm  \nonumber\\
                     & = \int(\pi^T B_R + \gamma^T M)f_{M|\Psi(X)=r}(M=m)dm  \nonumber\\
                     & = \pi^T B_R + \gamma^T E(M|\Psi(X)=r),  \nonumber
\end{flalign}
where $E(M|\Psi(X)=r) = (E(M_1|\Psi(X)=r), E(M_2|\Psi(X)=r), ..., E(M_m|\Psi(X)=r))^T$. 

The ITT treatment effect can then be computed by averaging the CATE over $\Psi$:
\begin{flalign}
\Delta_{ITT} & = \int_\Psi\Pi_{DB}(\Psi(X)=r)f(\Psi(X)=r)dr  \nonumber\\
             & = E_{\Psi}[\pi^T B_R + \gamma^T E(M|\Psi(X)=r)]  \nonumber\\
             & = \pi^T E_{\Psi}(B_R) + \gamma^T E(M).  \nonumber
\end{flalign}
If $B_R=(1,R,R^2,R^3)^T$, $E_{\Psi}(B_R) = (1,E(\Psi(X)), E(\Psi(X)^2), E(\Psi(X)^3))^T$, in which $E(\Psi(X))=E(E(\Delta_{DB} \mid A=0, X))=E(\Delta_{DB} \mid A=0)$, while $E(\Psi(X)^2)$ and $E(\Psi(X)^3)$ are the second and third moment of the $\Psi(X)$ distribution.

Similarly, for the first stage, the CATE function of placebo response and the $\Delta_{STD}$ can be expressed as:
\begin{flalign}
\Pi_{RW}(\Phi(X)=r) & = \pi^T B_R + \gamma^T E(M|\Phi(X)=r),  \nonumber \\
\Delta_{STD} & = \pi^T E_{\Phi}(B_R) + \gamma^T E(M).  \nonumber
\end{flalign}
And the difference is given by $\Delta_{ITT}-\Delta_{STD}= \pi^T[E_{\Psi}(B_R)-E_{\Phi}(B_R)]$. 
\end{proof}

Within this parametric framework, the theorem provides a simple method for estimating the standardized treatment effect $\Delta_{STD}$ from the observed $\Delta_{ITT}$. This requires accurate estimation of both the CATE parameters $\pi$ and the expectations of the basis functions $B_R$ under the distributions of the two prognostic scores, $\Psi(X)$ and $\Phi(X)$.

For illustration, if the CATE is a linear function of the placebo response $R$ (i.e., $\Pi(R,M) = \pi_0 + \pi_1 R + \gamma^\top M$), the underestimated treatment benefit given by this theorem is: $\Delta_{ITT} - \Delta_{STD} = \pi_1[E(\Delta_{DB} \mid A=0) - E(\Delta_{RW})]$. This expression can be estimated directly as the product of the slope parameter $\pi_1$ and the observed mean difference in placebo response between the double-blinded and the pragmatic single-blinded phases.

This derivation also clarifies a key nuance: when effect modifiers $M$ are present, the CATE functions $\Pi_{DB}(r)$ and $\Pi_{RW}(r)$ are not equivalent for a given placebo response level $r$. This is because the mean of $M$ conditional on the different prognostic scores are generally differ: $E(M|\Psi(X)=r) \neq E(M|\Psi(X)=r)]$.

Finally, we note that deriving closed-form analytical expressions for the CATE functions $\Pi_{ITT}(r)$ and $\Pi_{STD}(r)$ is complex. This complexity arises from the need to compute conditional expectations of the form $E(M|\Psi(X)=r)$ and $E(M|\Phi(X)=r)$. Without imposing stronger distributional assumptions (e.g., joint normality), these quantities generally do not simplify to tractable closed forms. 

In Appendix \ref{sec:sectionA}, we detail the computational derivations for a four-variable vector $X$ used in the low-dimensional simulation scenario. This framework assumes $(X_1, X_2, X_3)$ follow a multivariate normal distribution and $X_4$ is a binary variable independent of $(X_1, X_2, X_3)$. In Appendix \ref{sec:sectionB}, we present the derivation for a high-dimensional setting where $X$ is a $p$-dimensional multivariate normal random vector, with the first $m$ variables designated as effect modifiers $M$. This derivation underpins the high-dimensional simulation study. 


\section{Simulation}
\label{sec:Section4}
Simulation studies were performed to assess the statistical properties of the proposed methods for estimating $\Delta_{STD}$. Two data-generating scenarios, differing in dimensionality and functional form, were considered:

\begin{itemize}
    \item A low-dimensional scenario where the true prognostic score functions, $\Phi(X)$ and $\Psi(X)$, include only four baseline predictors. The CATE function, $\Pi(R,M)$, was modeled as a linear function of the placebo response $R$, with an additional two-dimensional effect modifier vector $M$.

    \item A high-dimensional setting includes 50 predictors in each prognostic score, where the CATE $\Pi(R,M)$ was defined as a non-linear function of $R$ that incorporated a varying number ($m = 2, 5, 10, 20$) of effect modifiers $M$.
\end{itemize}

The two simulation scenarios were designed with distinct objectives. The low-dimensional case was constructed for pedagogical clarity, allowing for a detailed examination of each step, including mathematical derivations, intermediate outputs, and trajectories of the CATE function after integration over $M$. This transparency allows us to validate the accuracy of each computational stage in a tractable setting. Conversely, the high-dimensional case emulates the properties of actual clinical trial data. Consequently, performance metrics such as the bias and mean squared error (MSE) derived from this scenario provide a realistic assessment of estimating $\Delta_{STD}$ in the practical data.

\subsection{The Data-Generating Process}
\label{sec:Section4.1}
\textbf{Scenario 1}\newline
To emulate the setting of a psychiatric trial evaluating a new medication for Major Depressive Disorder (MDD), a set of studies was simulated. The primary endpoint was the change in the Montgomery-Åsberg Depression Rating Scale (MADRS) score, a widely used clinician-administered measure of depression severity, in which lower scores correspond to a reduction in symptoms. The data-generating process includes the following steps:

Step 1: Randomly generate the treatment assignment indicator $A \sim Bernoulli(0.5)$, with $A=1$ denoting the active treatment arm and $A=0$ denoting the placebo arm.

Step 2: Generate the continuous baseline covariates $X=(X_1, X_2, X_3)^T \overset{\text{iid}}{\sim} MVN(\mu, \Sigma)$, where $\mu = (45, 0, 0)^T$ and $\Sigma = \begin{pmatrix} 16 & -2 & 1\\ -2 & 1 & 0.5\\ 1 & 0.5 & 1 \end{pmatrix}$. $X_1$ represents the baseline MADRS score, $X_2$ and $X_3$ represent two standardized normally distributed continuous variables with a correlation of 0.5. 

Step 3: Generate a binary baseline covariate $X_4$ independent of $X_1$ to $X_3$, with $X_4 \sim Bernoulli(0.7)$.

Step 4: Generate the expected potential outcomes (true prognostic scores) under placebo assignment during both the pragmatic first phase and the double-blinded phase, represented as $\Phi(X) = \alpha^T X = -15.6 + 0.25X_1 - X_3 + 0.5X_4$ and $\Psi(X) = \beta^T X = -44.45 + 0.75X_1 - 2X_2 + X_3 + X_4$, respectively. Consequently, the mean placebo response (mean reduction in MADRS) during the pragmatic stage is -4, while in the double-blinded phase, it is -10.

Step 5: Generate CATE $\Pi(R, M) = \Pi(\Psi(X)=\psi, M) = -5-0.2\psi+0.5X_2+0.4X_4$, in which $X_2$ and $X_4$ are effect modifiers $M$ in additional to the placebo response level $\psi$. 

Step 6: For each individual, generate the MADRS at the end of the first phase with an independent random error $\epsilon\sim N(\mu=0,\sigma^2=4)$, such that $Y_{RW} = X_1+\Phi(X) + \epsilon$. Next, generate the MADRS at the end of the double-blinded phase as $Y_{DB} = Y_{RW}+\Psi(X) +A*\Pi(\Psi(X), M) + \eta$, where $\eta\sim N(\mu=0,\sigma^2=4)$.

\begin{figure*}
    \centerline{\includegraphics[width=16cm, height=6cm]{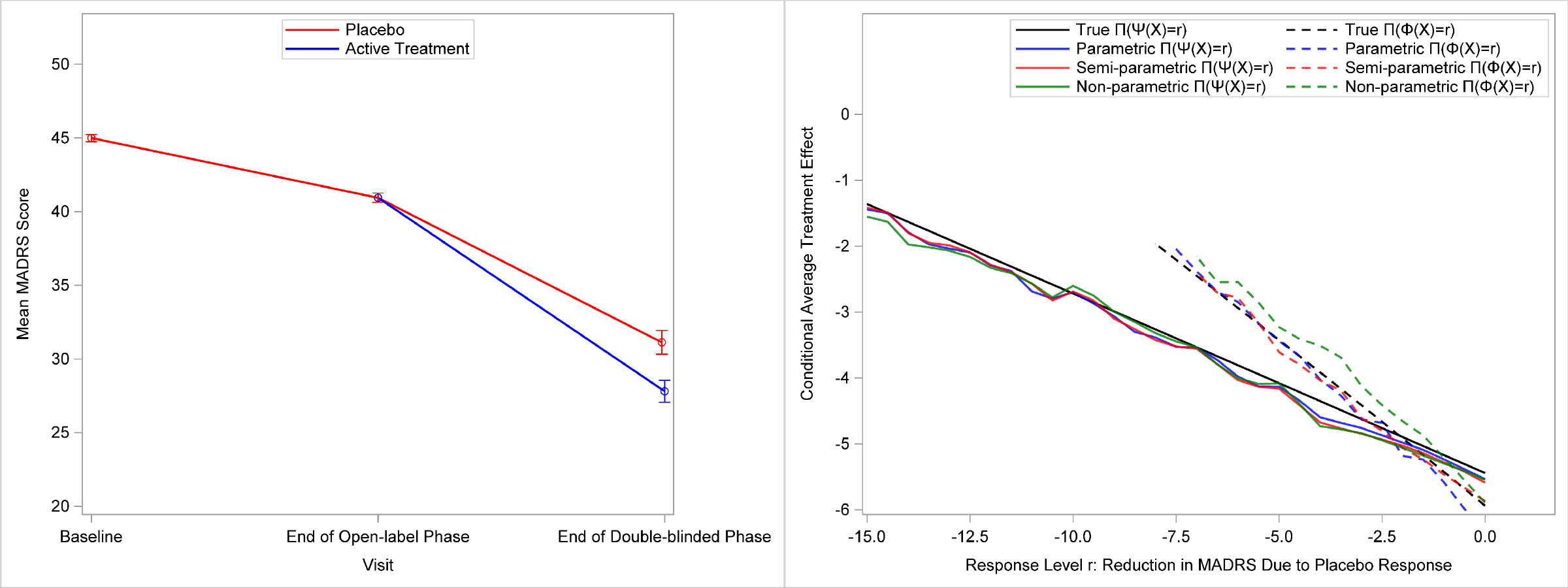}}
    \caption{Mean Trajectories and Estimation of CATE Functions}
    \label{fig:fig3}
\end{figure*}

Using one realization of simulated data with $N = 1000$ subjects, mean trajectory plots for both the active treatment group and the placebo arm are presented in Figure \ref{fig:fig3} Panel A, together with the theoretical and estimated CATE function $\Pi(R)$ (after integration over $M$) were shown in Figure \ref{fig:fig3}. The detailed derivation of the true CATE functions $\Pi_{DB}(\Psi(X)=r)$ and $\Pi_{RW}(\Phi(X)=r)$, alongside the theoretical values of $\Delta_{STD}$, is provided in Appendix \ref{sec:sectionA}. \newline

\noindent\textbf{Scenario 2}\newline
Scenario 2 followed similar steps to those in Scenario 1, with high-dimensional $X$ and $M$ used in the three functions $\Psi(X)$, $\Phi(X)$, and $\Pi(R,M)$. The detailed data-generating process is listed below:

Step 1: Randomly generate the treatment assignment indicator $A \sim Bernoulli(0.5)$, with $A=1$ denoting the active treatment arm and $A=0$ denoting the placebo arm.

Step 2: Generate the p-dimensional baseline predictors $X=(X_1, X_2, ..., X_p)^T \overset{\text{iid}}{\sim} MVN(\mu, \Sigma)$, where $\mu = 0_p^T$ and $\Sigma$ is a randomly generated $p \times p$ symmetric and positive definition variance-covariance matrix with diagonal elements of 1 and mild off diagonal correlations (between -0.3 and 0.3). 

Step 3: The true prognostic scores were set to $\Phi(X) = -4 + \alpha^T X$ and $\Psi(X) = -10 + \beta^T X$, with each element of the coefficient vectors $\alpha$ and $\beta$ independently sampled from a Uniform(-2, 2) distribution. This yields mean placebo responses of -4 and -10 for the pragmatic single-blinded placebo and double-blinded phases, respectively.

Step 4: The CATE was specified by the function: $\Pi(R, M) = \Pi(\Psi(X)=\psi, M) = -5-0.2\psi+0.01\psi^2+\gamma M$, where the effect modifier vector $M$ consists of the first $m$ components of $X$. 

Step 5: For each individual, generate the MADRS at the end of the first phase with an independent random error $\epsilon\sim N(\mu=0,\sigma^2=4)$, such that $Y_{RW} = Y_0 + \Phi(X) + \epsilon$, where $Y_0=40+2*X_1$. Next, generate the MADRS at the end of the double-blinded phase as $Y_{DB} = Y_{RW}+\Psi(X) +A*\Pi(\Psi(X), M) + \eta$, where $\eta\sim N(\mu=0,\sigma^2=4)$.

In Appendix \ref{sec:sectionB}, we have derived the theoretical value of $\Delta_{ITT}$ and $\Delta_{STD}$ under strong multivariate normal assumptions. \newline

Given the inherent complexity in estimating the prognostic scores $\Psi(X)$ and $\Phi(X)$ compared to the CATE function $\Pi(R,M)$, for both scenarios, we employed three estimation strategies to ensure a robust evaluation of $\Delta_{STD}$: 1) A full-parametric approach, using parametric linear regressions for all three functions. 2) A semi-parametric approach applying generalized additive models (GAMs) to $\Psi(X)$ and $\Phi(X)$ while retaining a parametric linear model for $\Pi(R,M)$. 3) A non-parametric approach, utilizing non-parametric GAMs for $\Psi(X)$, $\Phi(X)$, and $\Pi(R,M)$. 

In the first scenario, we evaluated the absolute bias, MSE, and coverage of the 95\% confidence intervals of the proposed method across sample sizes of SS = 100, 200, 500, 1000, 2000, and 5000. For each sample size and for parametric, semi-parametric, and non-parametric approaches, we generated 250 Monte Carlo replicates. Wald-type confidence intervals were constructed based on the 250 bootstrap resamples. To mitigate the risk of overfitting during the development of the prediction models of $\Psi(X)$, $\Phi(X)$, and $\Pi(R,M)$, we employed 5-fold cross-validation for parametric models and utilized generalized cross-validation to select optimal smoothing parameters for non-parametric models. For the high-dimensional setting in the second scenario, we evaluated absolute bias and MSE using 1,000 Monte Carlo replications, exclusively for larger samples (SS = 500, 1000, 2000, and 5000), as smaller sample sizes (SS = 100, 200) lacked sufficient degrees of freedom for non-parametric estimations. All procedures were implemented in SAS version 9.4, and the corresponding source code is available upon reasonable request.

\subsection{Simulation Results}
\label{sec:Section4.2}
\begin{figure*}
    \centerline{\includegraphics[width=16cm, height=8cm]{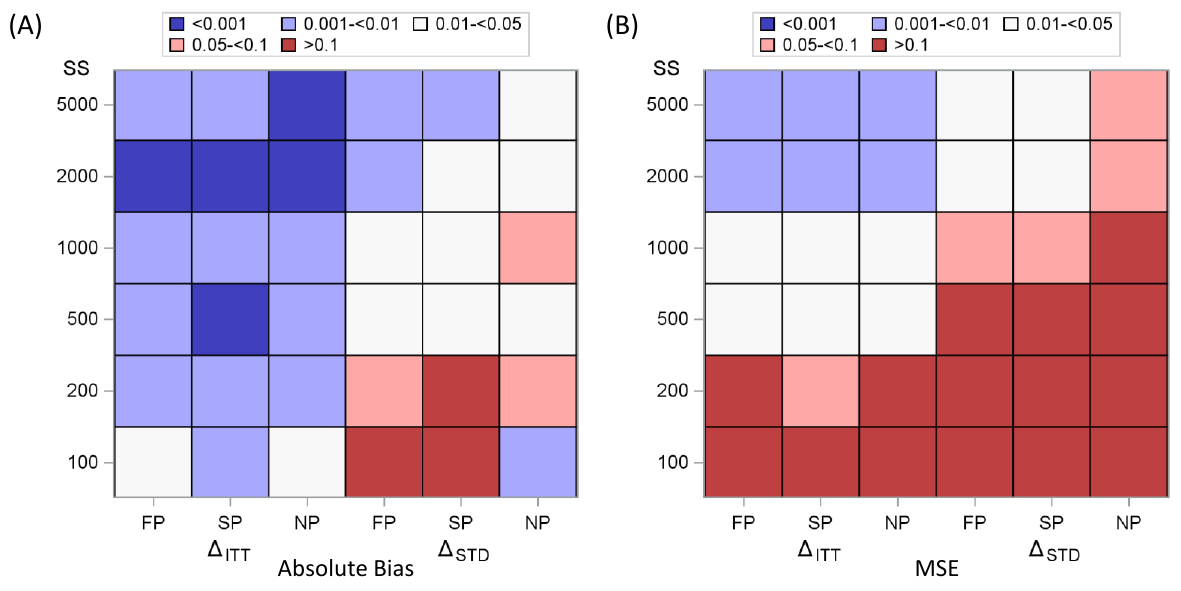}}
    \caption{Absolute Bias and MSE for Simulated Data in Scenario 1}
    \label{fig:fig4}
\end{figure*}

\noindent\textbf{Scenario 1}\newline
For this low-dimensional scenario, Table \ref{table:C1} and Figure \ref{fig:fig4} illustrate the absolute bias, MSE, and nominal coverage for the 95\% CI for both ITT effect $\Delta_{ITT}$ and the standardized treatment effect $\Delta_{STD}$. 

For $\Delta_{ITT}$, all three estimation methods exhibit minimal bias (close to 0) across all sample sizes, indicating that the ITT effect is robust and approximately unbiased for all three estimation approaches. The MSE decreases consistently as sample size increases for all methods, demonstrating the expected convergence. The 95\% confidence intervals generated using the bootstrap method consistently exhibit coverage probabilities near the nominal level (0.95) across all evaluated methods.

The $\Delta_{STD}$ shows notable bias at small sample sizes but decreases rapidly as the sample size grows. As expected, the MSE for $\Delta_{STD}$ is substantially higher than that of $\Delta_{ITT}$ for all three methods. However, MSE decreases sharply with increasing sample size for all methods. The full-parametric and semi-parametric methods outperform the non-parametric approach in terms of MSE across all sample sizes. The coverage probabilities for $\Delta_{STD}$ are generally near the nominal level (0.93 to 0.97).

A comparison of the estimation approaches revealed that the full-parametric and semi-parametric methods performed similarly for both estimators across all evaluated metrics (bias, MSE, coverage). Conversely, the non-parametric method demonstrated comparable performance only for the $\Delta_{ITT}$ estimator. Its performance in estimating $\Delta_{STD}$ was less efficient, particularly at small sample sizes where the MSE was significantly elevated. This finding is consistent with the known requirement for larger samples to stabilize non-parametric estimation $\Delta_{STD}$. \newline

\begin{figure*}
    \centerline{\includegraphics[width=16cm, height=9.6cm]{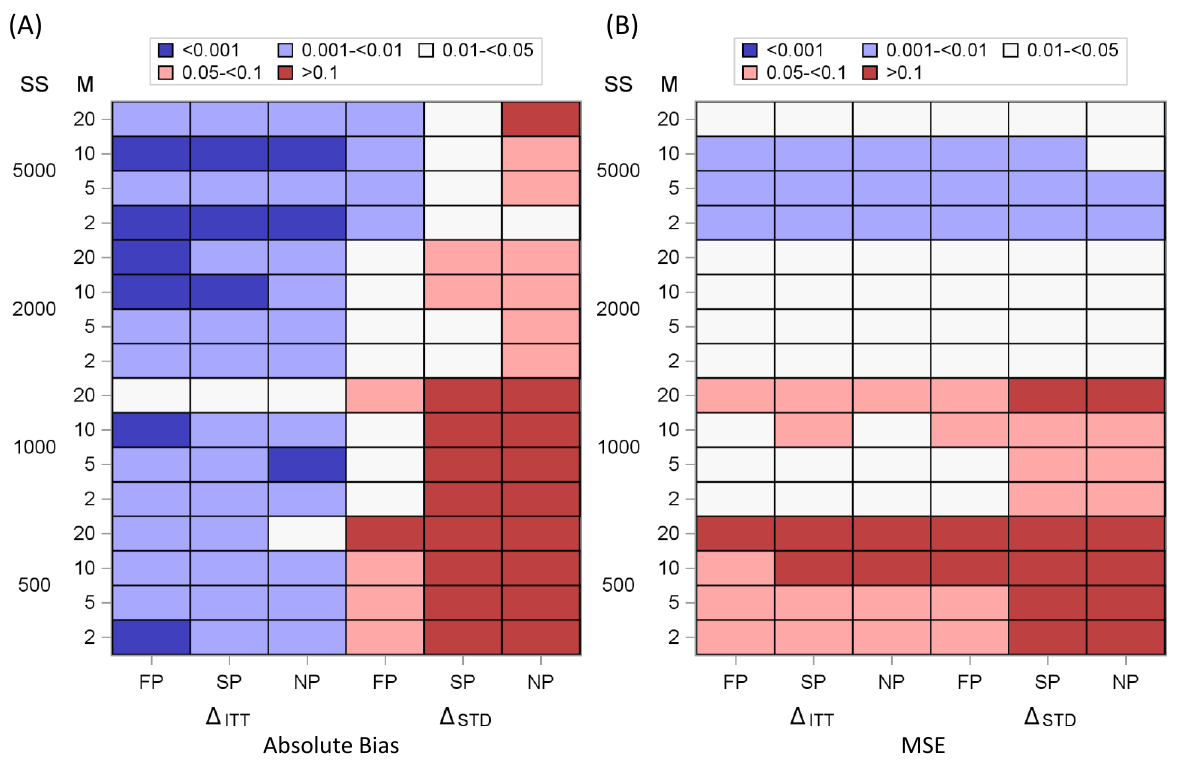}}
    \caption{Absolute Bias and MSE for Simulated Data in Scenario 2}
    \label{fig:fig5}
\end{figure*}

\noindent\textbf{Scenario 2}\newline
In this scenario, we examined the performance of estimators for $\Delta_{ITT}$ and $\Delta_{STD}$ under a high-dimensional setting, holding the number of prognostic score predictors constant at $p=50$. The simulation design varies three key factors: the sample sizes ($SS=500, 100, 200, 5000$), the dimension of the effect modifiers ($M = 2, 5, 10, 20$), and the class of estimation methodology (full-parametric, semi-parametric, and non-parametric). The numerical results for the simulated data are shown in Table \ref{table:C2}, and the corresponding heatmap is illustrated in Figure \ref{fig:fig5}.

The ITT effect $\Delta_{ITT}$ demonstrated remarkable robustness. Its absolute bias remained consistently low and unaffected by the number of effect modifiers across all estimation methods. While a modest increase in MSE was observed for higher dimensions of $M$, likely attributable to increased parametric variance, the performance of all three methods was virtually indistinguishable for estimating the ITT effect. 

The full-parametric method is, unsurprisingly, the most accurate and precise estimator for $\Delta_{STD}$. Its performance remains robust as the number of effect modifiers increases, with both absolute bias and MSE converging rapidly toward zero with larger sample sizes. For example, with $M=20$ and $SS=500$, it achieves a modest absolute bias of 0.1103 (2.8\% relative bias) and an MSE of 0.1771. When the sample size grows to $SS=5000$, these metrics improve dramatically to a near-negligible bias of 0.0056 and an MSE of just 0.0144. This confirms that under correct model specification, the full-parametric method is exceptionally reliable for estimating $\Delta_{STD}$.

As anticipated, the semi-parametric method, which relies on fewer model specifications and uses non-parametrically estimated prognostic scores, exhibits slightly inferior performance. This is especially evident with many effect modifiers ($M=20$) and a small sample size ($SS=500$), where its absolute bias of 0.3682 far exceeds the full-parametric method's 0.1103. However, the performance gap narrows considerably with more data. At $SS=5000$, the semi-parametric bias decreases to 0.0324, much closer to the full-parametric benchmark of 0.0056. Although the semi-parametric method has a consistently higher MSE than the full-parametric method, its precision becomes comparable when the number of effect modifiers is small ($M$=2 or 5).

The non-parametric method, which requires minimal model specification, has an MSE that is slightly higher yet very similar in magnitude to that of the semi-parametric method. However, its bias does not appear to approach zero as the sample size increases. For example, even with only two effect modifiers ($m=2$) and a large sample size ($N=5000$), the absolute bias is 0.0476. This is substantially higher than the absolute bias in the full-parametric method (0.0054) and the semi-parametric method (0.0196). This result demonstrates that while the MSE converges to zero (albeit at a slower rate), achieving an unbiased estimate numerically is difficult. Nevertheless, this bias is practically negligible for relatively large samples ($N \geq 2000$), resulting in a relative bias of around 1.5\%. Therefore, the method presents an acceptable and robust solution when model specification is infeasible.


\section{Discussion}
This article addresses the problem of underestimated treatment efficacy in trials with high placebo responses by introducing a two-stage framework to estimate a standardized causal effect, $\Delta_{STD}$. The design begins with a real-world pragmatic single-blinded placebo lead-in phase to control for contextual factors and mimic real-world medication use, followed by a double-blinded phase to estimate the CATE function. The resulting estimand, $\Delta_{STD}$, quantifies the treatment benefit for the ITT population by standardizing over the distribution of placebo responses measured in the initial phase. This approach provides a more realistic and generalizable measure of efficacy than the traditional trial-based ITT effect.

Building upon the methodological framework and contributions of this paper, several promising directions for future research naturally present themselves, including applied, theoretical, and new domain extensions. 

From an applied perspective, a primary next step involves rigorously validating the practical utility of the proposed framework by implementing it in new clinical studies with the expected high placebo responses, such as in psychiatry, neurology, and pain management. Subsequent work should also focus on enhancing the robustness of model specifications through a deeper investigation into the predictors underlying the prognostic scores and the effect modifiers within the CATE function. Such an inquiry would strengthen parametric and semi-parametric specifications, gain a better understanding of the direction and magnitude of the important predictors, and reduce the need to rely solely on non-parametric machine learning methods. Furthermore, the integration of Real-World Evidence (RWE) on predicting placebo responses could be highly valuable. Such external evidence could be systematically compared and validated against data from the pragmatic single-blinded placebo lead-in phase before integration. In extreme cases, a thoroughly validated external data source could potentially replace the first stage altogether and be used directly to augment the CATE function estimated from the double-blinded phase, thereby supporting the estimation of $\Delta_{STD}$.

On a more theoretical front, a few extensions are needed to broaden the applicability of this method. The current definition of the placebo response as an absolute change in a primary endpoint should be validated and compared with other secondary endpoints, and see which measure serves as a more sensitive and consistent measure, both internally across the two stages and externally across different studies. A significant challenge lies in generalizing the core theorem beyond effect measures on an absolute scale to relative scales, such as odds ratios and hazard ratios, and to complex settings, including survival analysis and repeated measures. To transition the framework from an analytical tool to a guide for trial design, future work must develop dedicated power calculation methods and adaptive designs that treat the standardized treatment benefit, $\Delta_{STD}$, as a primary efficacy measure. This would enable practical features, such as interim analyses for futility or sample size re-estimation based on the observed variation in placebo response. Finally, extending the methodology to non-inferiority trials or studies with multiple active comparators would require defining analogous standardized effects when the reference arm is an active treatment rather than a placebo.

An entirely new and critical direction involves adapting this methodology for medical device trials, where the ethical constraints against pragmatic single-blinded sham procedures present a unique challenge. In fields like cardiovascular research, estimating the (hypothetical) real-world sham response is impossible, necessitating novel approaches built on stronger assumptions. For instance, attributing a small portion of the improvement from the actual procedure to contextual factors. This extension, prompted by direct interest from the medical device industry, represents a complex but valuable frontier for this research. 

In sum, these future directions would not only solidify the methodological foundations laid out herein but also promote the broader adoption of causal inference methods for improving the design and interpretation of clinical trials when a high placebo/sham response is anticipated.

\section{Conclusion}
This two-stage framework provides clinical researchers and regulators with a principled, causal inference-based tool to isolate the pure pharmacological treatment effect from the high placebo response. A key advantage is that the design simultaneously yields two important causal estimates from the same ITT population: the traditional ITT effect ($\Delta_{ITT}$) from the double-blinded phase and the transported treatment effect ($\Delta_{STD}$). By quantifying the treatment effect relative to a realistic placebo response level, $\Delta_{STD}$ offers a more informative and generalizable measure of a medication's true efficacy for daily use.

Our work makes three key contributions. First, we provide a formal causal definition of the standardized treatment effect ($\Delta_{STD}$) as the average treatment effect standardized over the distribution of placebo responses estimated in the real-world pragmatic single-blinded placebo lead-in phase. Second, we establish a rigorous identification framework that leverages prognostic scores for dimension reduction, enabling the estimation of an interpretable CATE function, $\Pi(R, M)$, even with high-dimensional covariates. Third, we derive a theoretical theorem that quantifies the underestimated treatment benefit in the traditional ITT effect under parametric assumptions.

The simulation studies demonstrated the practical validity of our approach, showing that the proposed methods successfully recovered the true $\Delta_{STD}$ in both low and high-dimensional settings. Key findings confirm that $\Delta_{STD}$ is estimable with good precision, particularly at larger sample sizes. As anticipated, its mean squared error (MSE) was higher than that of $\Delta_{ITT}$, reflecting the increased complexity of the estimation process. Regarding methodological performance, the fully parametric approach proved to be the most accurate and efficient under model non-specification. Semi-parametric methods provided a robust alternative with only a minor loss in efficiency, while non-parametric methods, though unbiased for $\Delta_{ITT}$, required substantially larger samples to achieve consistent estimates for $\Delta_{STD}$.

\section*{Conflict of interest}
The authors do not have any personal financial interests related to the subject matters discussed in this manuscript.

\bibliographystyle{unsrt}
\bibliography{wileyNJD-AMA}

\newpage
\appendix
\section{Derivation of the Theoretical Values for Scenario 1}
\label{sec:sectionA}
In this scenario 1, the theorem provides us the opportunity to calculate the theoretical values for $\Pi_{DB}(\Psi(X)=r)$, $\Delta_{ITT}$, $\Pi_{RW}(\Phi(X)=r)$, and $\Delta_{STD}$.

To derive the true CATE function for $\Pi_{DB}(\Psi(X)=r)=\pi^TB_R+\gamma^TE(M \mid \Psi(X)=r)$, we need to find $E(M \mid \Psi(X)=r) = (E(X_2 \mid \Psi(X)=r), E(X_4 \mid \Psi(X)=r))^T$. Let $X=(X_1,X_2,X_3)^T$, and $\beta=(\beta_1,\beta_2,\beta_3)^T$, such that $\Psi(X)=\beta_0+\beta^TX+\beta_4X_4$, $E(X_4 \mid \Psi(X)=r)$ can be calculated using Bayes' theorem as:
\begin{flalign}
E(X_4 \mid \Psi(X)=r) & = \frac{f(\Psi(X)=r \mid X_4=1) P(X_4=1)}{f(\Psi(X)=r \mid X_4=1) P(X_4=1) + f(\Psi(X)=r \mid X_4=0) P(X_4=0)} \nonumber\\
                     & = \frac{f(\beta^TX=r-\beta_0-\beta_4 \mid X_4=1) P(X_4=1)}{f(\beta^TX=r-\beta_0-\beta_4 \mid X_4=1) P(X_4=1) + f(\beta^TX=r-\beta_0 \mid X_4=0) P(X_4=0)}  \nonumber\\
                     & = \frac{f(\beta^TX=r-\beta_0-\beta_4) P(X_4=1)}{f(\beta^TX=r-\beta_0-\beta_4) P(X_4=1) + f(\beta^TX=r-\beta_0) P(X_4=0)}  \nonumber\\
                     & = \frac{\phi\left(\frac{r-\beta_0-\beta_4-\beta^T\mu}{\sqrt{\beta^T\Sigma\beta}}\right) P(X_4=1)}{\phi\left(\frac{r-\beta_0-\beta_4-\beta^T\mu}{\sqrt{\beta^T\Sigma\beta}}\right) P(X_4=1) + \phi\left(\frac{r-\beta_0-\beta^T\mu}{\sqrt{\beta^T\Sigma\beta}}\right) P(X_4=0)}  \nonumber
\end{flalign}
In which $\phi$ is the PDF for the standard normal distribution, as the linear combination of the multivariate normal covariates $X$ is normally distributed, that is, $\beta^TX \sim N(\beta^T\mu, \beta^T\Sigma\beta)$. To calculate $E(X_2|\Psi(X)=r)$:
\begin{flalign}
E(X_2 \mid \Psi(X)=r) & = E(X_2 \mid \Psi(X)=r, X_4=1)P(X_4=1 \mid \Psi(X)=r) + E(X_2 \mid \Psi(X)=r, X_4=0)P(X_4=0 \mid \Psi(X)=r) \nonumber \\
                      & = E(X_2 \mid \beta^TX=r-\beta_0-\beta_4)P(X_4=1 \mid \Psi(X)=r) + E(X_2 \mid \beta^TX=r-\beta_0)P(X_4=0 \mid \Psi(X)=r), \nonumber
\end{flalign}
where $P(X_4=1 \mid \Psi(X)=r) = E(X_4=1 \mid \Psi(X)=r)$, and this term has just been derived. To find the conditional mean of $X_2|\beta^TX$, let the vector $C_2 = (0, 1, 0)^T$, then the conditional distribution of the multivariate normal variables gives us:
\begin{flalign}
\begin{pmatrix} X_2 \\ \beta^TX \end{pmatrix} {\sim} N\Bigg(\begin{pmatrix} \mu_2 \\ \beta^T\mu \end{pmatrix}, \begin{pmatrix} \sigma^2_{22} & C^T_2 \Sigma \beta \\ C^T_2 \Sigma \beta & \beta^T \Sigma \beta \end{pmatrix}\Bigg) \Rightarrow X_2 \mid \beta^TX \sim N(\mu_{X_2 \mid \beta^TX}, \sigma^2_{X_2 \mid \beta^TX}), \nonumber
\end{flalign}
where $\mu_{X_2 \mid \beta^TX} = \mu_2 + (C^T_2 \Sigma \beta)(\beta^T \Sigma \beta)^{-1}(\beta^TX-\beta^T\mu)$, and $\sigma^2_{X_2 \mid \beta^TX} = \sigma^2_{22} - (C^T_2 \Sigma \beta)(\beta^T \Sigma \beta)^{-1}(C^T_2 \Sigma \beta)$. Therefore, $E(X_2|\beta^TX=r-\beta_0-\beta_4) = \mu_2 + (C^T_2 \Sigma \beta)(\beta^T \Sigma \beta)^{-1}(r-\beta_0-\beta_4-\beta^T\mu)$, and $E(X_2|\beta^TX=r-\beta_0) = \mu_2 + (C^T_2 \Sigma \beta)(\beta^T \Sigma \beta)^{-1}(r-\beta_0-\beta^T\mu)$.

Plug-in numbers and we obtain the true CATE function $\Pi_{DB}(\Psi(X)=r) = -5.823 - 0.277r + \frac{0.477}{1+exp(-0.051r-0.523)}$ for the double-blinded phase. For the pragmatic first phase, a similar process to replace $\beta$ using $\alpha$ would give the true CATE function $\Pi_{RW}(\Phi(X)=r) = -6.450 - 0.533r + \frac{0.567}{1+exp(-0.333r-1.367)}$. 

For true value of the marginal treatment effects, $\Delta_{ITT} = \pi^T E_{\Psi}(B_R) + \gamma^T E(M) = -5-0.2*(-10) + 0.5*(0) + 0.4*0.7 = -2.72$, and $\Delta_{STD} = \Delta_{ITT} - \pi^T[E_\Psi(B_R)-E_\Phi(B_R)] = -2.72 + 0.2(-10 + 4) = -3.92$.

\newpage
\section{Theoretical Derivation Under Parametric and Normality Assumptions}
\label{sec:sectionB}
In this particular scenario, alongside the assumptions B1 and B2 detailed in section \ref{sec:Section3.4}, we present additional assumptions outlined below.
\begin{itemize}
\item B3: Let the p-dimensional random vector of the baseline covariates $X = (X_1, X_2, ..., X_P)^T \overset{\text{iid}}{\sim} N(\mu_X, \Sigma_X)$, in which $X_1$ captures the baseline severity score $Y_0$. Without loss of generality, we can standardize each random variable $X_i$, resulting in the standardized covariates $Z = D^{-\frac{1}{2}}(X-\mu_X) \overset{\text{iid}}{\sim} N(0, \Sigma)$, where $D = diag(\Sigma_X)$ and $\Sigma=D^{-\frac{1}{2}}\Sigma_XD^{-\frac{1}{2}}$, with the diagonal elements of $\Sigma$ equal to 1.  

\item B4: Assume both prognostic scores are linear functions of baseline covariates, expressed as $\Phi(Z) = \alpha^T Z+\lambda_{RW}$ for the real-world, pragmatic first phase, and $\Psi(Z) = \beta^T Z+\lambda_{DB}$ for the double-blinded phase. For the standardized random variables $Z$, $\lambda_{RW}=E(\Delta_{RW})$ represents the mean placebo response observed during the first stage, while $\lambda_{DB}=E(\Delta_{DB} \mid A=0)$ quantifies the mean placebo response during the double-blinded stage.
\end{itemize}

Given the assumptions B1 to B4, the conditional average treatment effect functions $\Pi_{ITT}(\Psi(Z)=r)$ can be derived as follows:
\begin{flalign}
\Pi_{ITT}(\Psi(Z)=r) & = \int\Pi(\Psi(Z)=r,M=m)f_{M|\Psi(Z)=r}(M=m)dm  \nonumber\\
                     & = \int(\pi^T B_R + \gamma^T M)f_{M|\Psi(Z)=r}(M=m)dm  \nonumber\\
                     & = \pi^T B_R + \gamma^T E(M|\Psi(Z)=r),  \nonumber
\end{flalign}

in which, $E(M|\Psi(Z)=r) = (E(M_1|\Psi(Z)=r), E(M_2|\Psi(Z)=r), ..., E(M_m|\Psi(Z)=r))^T$. 

Let $C_M$ be a $p \times m$ matrix that chooses the effect modifiers vector $M$ from vector $Z$. Without loss of generality, by switching rows, assume the first $m$ covariates in $Z$ are effect modifiers, that is, $M=(Z_1,Z_2,...,Z_m)^T$. Then $C_M^T = (I_{m\times m},0_{m \times (p-m)})$, such that $C_M^TZ=(M,0_{1\times (p-m)}))^T$. The joint multivariate normal distribution of $(M, \Psi(Z))$ can be expressed as:

\begin{flalign}
\begin{pmatrix} M \\ \Psi(Z) \end{pmatrix} {\sim} N\Bigg(\begin{pmatrix} 0 \\ \lambda_{DB} \end{pmatrix}, \begin{pmatrix} C^T_M \Sigma C_M & C^T_M \Sigma \beta \\ C^T_M \Sigma \beta & \beta^T \Sigma \beta \end{pmatrix}\Bigg) \nonumber
\end{flalign}

Consequently, the conditional distribution $M|\Psi(Z)=r$ can be derived from multivariate normal theory:
\begin{flalign}
M|\Psi(Z)=r {\sim} N\Bigg(\frac{C^T_M \Sigma \beta}{\beta^T \Sigma \beta}(r-\lambda_{DB}), C^T_M \Sigma C_M - (C^T_M \Sigma \beta)(\beta^T \Sigma \beta)^{-1}(C^T_M \Sigma \beta)^T \Bigg). \nonumber
\end{flalign}

Therefore, the CATE function $\Pi_{ITT}(\Psi(Z)=r)$ is:
\begin{flalign}
\Pi_{ITT}(\Psi(Z)=r) & = \pi^T B_R + \gamma^T E(M|\Psi(Z)=r),  \nonumber \\
                     & = \pi^T B_R + \gamma^T \frac{C^T_M \Sigma \beta}{\beta^T \Sigma \beta}(r-\lambda_{DB}).  \nonumber
\end{flalign}

Similarly, the transported CATE function $\Pi_{STD}(\Phi(Z)=r)$ for real-world stage is calculated as: $\Pi_{STD}(\Phi(Z)=r) = \pi^T B_R + \gamma^T \frac{C^T_M \Sigma \alpha}{\alpha^T \Sigma \alpha}(r-\lambda_{RW})$.

When using standardized covariates with means of 0, the marginal ITT and standardized causal treatment effect can be written as:
\begin{flalign}
\Delta_{ITT} & = \int_\Psi\Pi_{ITT}(\Psi(X)=r)f(\Psi(X)=r)dr  \nonumber\\
             & = E_{\Psi}[\pi^T B_R + \gamma^T E(M|\Psi(X)=r)]  \nonumber\\
             & = \pi^T E_{\Psi}(B_R) + \gamma^T E(M)  \nonumber\\
             & = \pi^T E_{\Psi}(B_R),  \nonumber\\
\Delta_{STD} & = \pi^T E_{\Phi}(B_R).  \nonumber
\end{flalign}

For example, if the CATE is modeled as a linear function of the placebo response level $R$, where $\Pi(R,M) = \pi_0 + \pi_1R + \gamma^TM$. We can derive the ITT treatment effect as $\Delta_{ITT}=\pi_0 + \pi_1\lambda_{DB}$. Similarly, the standardized treatment effect is given by $\Delta_{STD}=\pi_0 + \pi_1\lambda_{RW}$, with the difference expressed as $\pi_1(\lambda_{DB}-\lambda_{RW})$. In this scenario, the sample mean of improvement observed at the double-blinded phase ($\Bar{\Delta}_{DB} \mid A=0$) and real-world, pragmatic, single-blinded placebo lead-in phase ($\Bar{\Delta}_{RW}$) serve as good estimates of $\lambda_{DB}$ and $\lambda_{RW}$, respectively. Hence, the estimation of $\pi_1$ becomes the crucial element in this analysis.

\newpage
\section{Detailed Simulation Results for Scenario 1 and 2}
\label{sec:sectionC}
\begin{table}[!ht]
    \centering
    \caption{Simulation Results for $\Delta_{ITT}$ and $\Delta_{STD}$ In Scenario 1}
    \begin{tabular}{|c|c|c|c|c|c|c|}
    \hline
        \textbf{} & \multicolumn{3}{c|}{$\Delta_{ITT}$} & \multicolumn{3}{c|}{$\Delta_{STD}$} \\ \hline
        \textbf{Sample Size} & \textbf{Abs Bias} & \textbf{MSE} & \textbf{Coverage} & \textbf{Abs Bias} & \textbf{MSE} & \textbf{Coverage} \\ \hline
        \multicolumn{7}{|c|}{\textbf{Full-parametric}} \\ \hline
        100 & 0.031 & 0.181 & 0.964 & 0.243 & 0.877 & 0.960 \\ \hline
        200 & 0.001 & 0.111 & 0.904 & 0.055 & 0.353 & 0.964 \\ \hline
        500 & 0.002 & 0.038 & 0.924 & 0.040 & 0.128 & 0.936 \\ \hline
        1000 & 0.003 & 0.018 & 0.944 & 0.015 & 0.063 & 0.952 \\ \hline
        2000 & 0.001 & 0.009 & 0.940 & 0.006 & 0.031 & 0.932 \\ \hline
        5000 & 0.002 & 0.004 & 0.960 & 0.003 & 0.012 & 0.932 \\ \hline
        \multicolumn{7}{|c|}{\textbf{Semi-parametric}} \\ \hline
        100 & 0.005 & 0.190 & 0.952 & 0.286 & 0.865 & 0.961 \\ \hline
        200 & 0.004 & 0.093 & 0.935 & 0.123 & 0.350 & 0.956 \\ \hline
        500 & 0.000 & 0.036 & 0.931 & 0.046 & 0.132 & 0.941 \\ \hline
        1000 & 0.006 & 0.018 & 0.945 & 0.021 & 0.061 & 0.947 \\ \hline
        2000 & 0.001 & 0.009 & 0.948 & 0.013 & 0.031 & 0.941 \\ \hline
        5000 & 0.004 & 0.003 & 0.957 & 0.007 & 0.013 & 0.934 \\ \hline
        \multicolumn{7}{|c|}{\textbf{Non-parametric}} \\ \hline
        100 & 0.033 & 0.176 & 0.968 & 0.005 & 2.096 & 0.972 \\ \hline
        200 & 0.006 & 0.113 & 0.912 & 0.067 & 0.932 & 0.948 \\ \hline
        500 & 0.004 & 0.039 & 0.932 & 0.013 & 0.307 & 0.960 \\ \hline
        1000 & 0.002 & 0.019 & 0.936 & 0.057 & 0.158 & 0.940 \\ \hline
        2000 & 0.000 & 0.009 & 0.944 & 0.035 & 0.080 & 0.920 \\ \hline
        5000 & 0.000 & 0.006 & 0.942 & 0.010 & 0.060 & 0.930 \\ \hline
    \end{tabular}
    \label{table:C1}
\end{table}

\begin{table}[!ht]
    \centering
    \caption{Simulation Results for $\Delta_{ITT}$ and $\Delta_{STD}$ In Scenario 2}
    \resizebox{\textwidth}{!}{
    \begin{tabular}{|c|c|c|c|c|c|c|c|c|c|c|c|c|}
    \hline
        \textbf{} & \multicolumn{4}{c|}{\textbf{Full-parametric}} & \multicolumn{4}{c|}{\textbf{Semi-parametric}} & \multicolumn{4}{c|}{\textbf{Non-parametric}} \\ \hline
        \textbf{} & \multicolumn{2}{c|}{$\Delta_{ITT}$} & \multicolumn{2}{c|}{$\Delta_{STD}$} & \multicolumn{2}{c|}{$\Delta_{ITT}$} & \multicolumn{2}{c|}{$\Delta_{STD}$} & \multicolumn{2}{c|}{$\Delta_{ITT}$} & \multicolumn{2}{c|}{$\Delta_{STD}$} \\ \hline
        \textbf{Sample Size} & Abs Bias & MSE & Abs Bias & MSE & Abs Bias & MSE & Abs Bias & MSE & Abs Bias & MSE & Abs Bias & MSE \\ \hline
        \multicolumn{13}{|c|}{\textbf{Dimension of M = 2}} \\ \hline
        500 & 0.0004 & 0.0692 & 0.0630 & 0.0774 & 0.0061 & 0.0831 & 0.2249 & 0.1446 & 0.0079 & 0.0824 & 0.2573 & 0.1584 \\ \hline
        1000 & 0.0027 & 0.0327 & 0.0326 & 0.0345 & 0.0045 & 0.0356 & 0.1057 & 0.0512 & 0.0043 & 0.0355 & 0.1237 & 0.0551 \\ \hline
        2000 & 0.0022 & 0.0147 & 0.0103 & 0.0153 & 0.0024 & 0.0153 & 0.0430 & 0.0182 & 0.0028 & 0.0152 & 0.0645 & 0.0209 \\ \hline
        5000 & 0.0003 & 0.0062 & 0.0054 & 0.0061 & 0.0004 & 0.0064 & 0.0196 & 0.0068 & 0.0001 & 0.0062 & 0.0476 & 0.0087 \\ \hline
        \multicolumn{13}{|c|}{\textbf{Dimension of M = 5}} \\ \hline
        500 & 0.0061 & 0.0700 & 0.0669 & 0.0701 & 0.0022 & 0.0838 & 0.2290 & 0.1396 & 0.0016 & 0.0833 & 0.2625 & 0.1590 \\ \hline
        1000 & 0.0036 & 0.0339 & 0.0322 & 0.0344 & 0.0011 & 0.0361 & 0.1039 & 0.0505 & 0.0003 & 0.0361 & 0.1256 & 0.0562 \\ \hline
        2000 & 0.0062 & 0.0160 & 0.0148 & 0.0156 & 0.0060 & 0.0165 & 0.0472 & 0.0188 & 0.0058 & 0.0165 & 0.0689 & 0.0219 \\ \hline
        5000 & 0.0023 & 0.0064 & 0.0051 & 0.0060 & 0.0021 & 0.0065 & 0.0194 & 0.0065 & 0.0021 & 0.0064 & 0.0514 & 0.0092 \\ \hline
        \multicolumn{13}{|c|}{\textbf{Dimension of M = 10}} \\ \hline
        500 & 0.0090 & 0.0938 & 0.0674 & 0.1133 & 0.0040 & 0.1064 & 0.2612 & 0.1978 & 0.0023 & 0.1049 & 0.2914 & 0.2123 \\ \hline
        1000 & 0.0005 & 0.0481 & 0.0328 & 0.0514 & 0.0014 & 0.0504 & 0.1217 & 0.0693 & 0.0025 & 0.0497 & 0.1379 & 0.0746 \\ \hline
        2000 & 0.0008 & 0.0221 & 0.0121 & 0.0245 & 0.0004 & 0.0225 & 0.0537 & 0.0279 & 0.0013 & 0.0223 & 0.0689 & 0.0314 \\ \hline
        5000 & 0.0007 & 0.0092 & 0.0021 & 0.0091 & 0.0002 & 0.0091 & 0.0197 & 0.0098 & 0.0000 & 0.0090 & 0.0594 & 0.0129 \\ \hline
        \multicolumn{13}{|c|}{\textbf{Dimension of M = 20}} \\ \hline
        500 & 0.0016 & 0.1294 & 0.1103 & 0.1771 & 0.0063 & 0.1398 & 0.3682 & 0.3228 & 0.0137 & 0.1433 & 0.3825 & 0.3303 \\ \hline
        1000 & 0.0108 & 0.0665 & 0.0632 & 0.0807 & 0.0118 & 0.0680 & 0.1878 & 0.1170 & 0.0130 & 0.0697 & 0.2021 & 0.1218 \\ \hline
        2000 & 0.0007 & 0.0325 & 0.0296 & 0.0361 & 0.0018 & 0.0334 & 0.0884 & 0.0453 & 0.0013 & 0.0340 & 0.0973 & 0.0494 \\ \hline
        5000 & 0.0034 & 0.0136 & 0.0056 & 0.0144 & 0.0037 & 0.0137 & 0.0324 & 0.0158 & 0.0032 & 0.0139 & 0.1085 & 0.0269 \\ \hline
    \end{tabular}
    }
    \label{table:C2}
\end{table}

\newpage
\section{When $M$ and $R$ Interacts in the CATE Function $\Pi(R,M)$}
\label{sec:sectionD}
Section \ref{sec:Section4.2} presented simulation studies validating the proposed methods for estimating the transported causal treatment effect, $\Delta_{STD}$, under the assumption from the quantification of the underestimated treatment benefit theorem that parameters $R$ and $M$ do not interact within the CATE function $\Pi(R,M)$. In this appendix, we extend the framework to the more complex scenario where $R$ and $M$ do interact, demonstrating through theoretical derivations and simulations that the proposed methods remain applicable.

In Section \ref{sec:SectionD1}, we generalize the Theorem to allow for a higher-order effect modification of $R$ by $M$. This generalization introduces analytical complexity, and a closed-form solution for the theoretical value of $\Delta_{STD}$ is generally unavailable. Subsequently, in Section \ref{sec:SectionD2}, we modify the data generation model to include this interaction term and generate datasets with sample sizes of N = 1,000 and N = 5,000. Using the identification and estimation approaches detailed in Sections X and Y, we estimate $\Delta_{STD}$ parametrically, semi-parametrically, and non-parametrically across varying strengths of the second-order effect modifier. The results confirm that the estimates align with theoretical expectations, exhibiting low absolute bias and MSE. Although the framework can scientifically accommodate such higher-order interactions, in Section \ref{sec:SectionD3}, we provide justifications in the real-data analysis for why the higher-order effect modifiers are less likely exist in data and often times not considered in practices. 

\subsection{Theoretical Derivation When $M$ and $R$ Interacts}
\label{sec:SectionD1}
In this subsection, we retain Assumption B1 and relax Assumption B2 to extend the Theorem in section \ref{sec:Section3.4}, allowing for higher-order effect modification by $M$ on $R$. The revised Assumption B2 explicitly accounts for interactions between the basis functions of $R$ and the other effect modifiers $M$.

\begin{itemize}
\item Revised Assumption B2: Assume that the CATE function is linear in the basis expansion of the placebo response $R$ and other effect modifiers $M$, such that $\Pi(R,M) = \pi^T B_R + \gamma^T M + \lambda^T B_R \otimes M$. Here, $B_R$ is a p-dimensional basis function for $R$, and $M$ is a m-dimensional vector of effect modifiers, and the parameter vectors $\pi$, $\gamma$, and $\lambda$ are $p$, $m$, and $p*m$ dimensions, respectively. 
\end{itemize}

This formulation explicitly account for the higher-order interaction terms between $R$ and $M$ in the Kronecker product $B_R \otimes M$. For example, let $B_R = (R,R^2)^T$ and $M=(M_1,M_2,M_3)^T$, the Kronecker product can be viewed as a form of vectorization of the outer product, which is, $B_R \otimes M=(R*M_1,R*M_2,R*M_3,R^2*M_1,R^2*M_2,R^2*M_3)^T$. \newline

\noindent\textbf{Theorem (Quantification of the Underestimated Treatment Benefit When $M$ and $R$ Interacts)}\newline
Given the assumptions B1 and revised assumption B2, the underestimated treatment benefit due to excessive placebo response in the double-blinded phase of using $\Delta_{ITT}$ as study treatment effect is given by:
\begin{flalign}
\Delta_{ITT}-\Delta_{STD} = & \pi^T[E_{\Psi}(B_R)-E_{\Phi}(B_R)] \nonumber \\ 
                        & + \lambda^T[E_{\Psi}(B_R \otimes E(M|\Psi(X)) - E_{\Phi}(B_R \otimes E(M|\Phi(X))]  \nonumber
\end{flalign}
where $E_{\Phi}(B_R)$ denotes the expectation of $B_R$ with respect to the distribution of the prognostic score $\Phi(X)$ from the pragmatic single-blinded phase, and $E_{\Psi}(B_R)$ denotes the expectation with respect to the distribution of $\Psi(X)$ from the double-blinded phase. The term $E_{\Psi}(B_R \otimes E(M|\Psi(X))$ denotes the expectation of outer product of $B_R$ and $E(M|\Psi(X))$ (which is also a function of response level $R$) with respect to the distribution of $\Psi(X)$. Similarly, the term $E_{\Phi}(B_R \otimes E(M|\Phi(X))$ denotes the expectation of the outer product with respect to the distribution of $\Phi(X)$.

\begin{proof}
Let $M=(M_1, M_2, ..., M_m)^T$ be the vector of effect modifiers. For the double-blinded stage, the CATE as a function of placebo response level can be derived as:
\begin{flalign}
\Pi_{DB}(\Psi(X)=r) & = \int\Pi(\Psi(X)=r,M=m)f_{M|\Psi(X)=r}(M=m)dm  \nonumber\\
                     & = \int(\pi^T B_R + \gamma^T M + \lambda^T B_R \otimes M)f_{M|\Psi(X)=r}(M=m)dm  \nonumber\\
                     & = \pi^T B_R + \gamma^T E(M|\Psi(X)=r) + \lambda^T E(B_R \otimes M|\Psi(X)=r),  \nonumber
\end{flalign}
where $E(M|\Psi(X)=r) = (E(M_1|\Psi(X)=r), E(M_2|\Psi(X)=r), ..., E(M_m|\Psi(X)=r))^T$. When $\Psi(X)=r$, $B_R$ as a vector function of $r$ are known and constant, and therefore, $\lambda^T E(B_R \otimes M|\Psi(X)=r) = \lambda^TB_R \otimes E(M|\Psi(X)=r)$.

The ITT treatment effect can then be computed by averaging the CATE over $\Psi$:
\begin{flalign}
\Delta_{ITT} & = \int_\Psi\Pi_{DB}(\Psi(X)=r)f(\Psi(X)=r)dr  \nonumber\\
             & = E_{\Psi}[\pi^T B_R + \gamma^T E(M|\Psi(X)=r) + \lambda^TB_R \otimes E(M|\Psi(X)=r)]  \nonumber\\
             & = \pi^T E_{\Psi}(B_R) + \gamma^T E(M) + \lambda^T E_{\Psi}[B_R \otimes E(M|\Psi(X)=r)].  \nonumber
\end{flalign}

As $R$ and $M$ are not independent, because some effect modifiers included in $M$ might also be a predictor included in the prognostic score functions in predicting the placebo response levels $R$, the expectation $E_{\Psi}[B_R \otimes E(M|\Psi(X)=r)]$ cannot be further simplified. 

Similarly, for the pragmatic single-blinded placebo phase, the CATE function of placebo response and the $\Delta_{STD}$ can be expressed as:
\begin{flalign}
\Pi_{RW}(\Phi(X)=r) & = \pi^T B_R + \gamma^T E(M|\Phi(X)=r) + \lambda^T E(B_R \otimes M|\Phi(X)=r),  \nonumber \\
\Delta_{STD} & = \pi^T E_{\Phi}(B_R) + \gamma^T E(M) + \lambda^T E_{\Phi}[B_R \otimes E(M|\Phi(X)=r)].  \nonumber
\end{flalign}
And the difference is given by $\Delta_{ITT}-\Delta_{STD}= \pi^T[E_{\Psi}(B_R)-E_{\Phi}(B_R)] + \lambda^T[E_{\Psi}(B_R \otimes E(M|\Psi(X)) - E_{\Phi}(B_R \otimes E(M|\Phi(X))]$. 
\end{proof}

Due to the absence of a general closed-form formula for $E[B_R \otimes E(M|R)]$, we resorted to Monte Carlo simulation to calculate its theoretical values in the next subsection.

\subsection{Simulation Studies When $M$ and $R$ Interacts}
\label{sec:SectionD2}
This subsection extends Simulation Scenario 1 by incorporating an additional second-order effect modifier. We evaluate the performance of the proposed identification method using absolute Bias and MSE by varying the strength of this interaction term across two sample sizes (N = 1,000 and N = 5,000).\newline

\noindent\textbf{The Data-Generating Process}\newline
The data-generating process follows the steps outlined in Section \ref{sec:Section4.1} (Scenario 1), with a simple modification to the CATE function in Step 5 specified below. To introduce a second-order interaction, we add the term $\psi*X_4$, where the parameter $\lambda$ controls the strength of this effect modification. 

\begin{itemize}
\item Revised Step 5: Generate CATE $\Pi(R, M) = \Pi(\Psi(X)=\psi, M) = -5-0.2\psi+0.5X_2+0.4X_4+\lambda \psi*X_4$, in which $X_2$ and $X_4$ are effect modifiers $M$ in additional to the placebo response level $\psi$. 
\end{itemize}

We varied the parameter $\lambda$ from 0.02 to 0.2 in increments of 0.02. For each value of $\lambda$, the theoretical values of $\Delta_{ITT}$ and $\Delta_{STD}$ were calculated and presented below in Table \ref{table:D1}.

\begin{table}[!ht]
    \renewcommand{\thetable}{D1}
    \centering
    \caption{\textbf{Theoretical Values for $\Delta_{ITT}$ and $\Delta_{STD}$ In the Revised Scenario 1 When $R$ and $M$ Interacts}}
    \renewcommand{\arraystretch}{1.0}
    \begin{tabular}{|c|c|c|}
    \hline
        \textbf{} $\lambda$ & $\Delta_{ITT}$ & $\Delta_{STD}$ \\ \hline
                    0.02   & -2.8558    & -3.9739    \\ \hline
                    0.04   & -2.9916    & -4.0278    \\ \hline
                    0.06   & -3.1274    & -4.0817    \\ \hline
                    0.08   & -3.2632    & -4.1356    \\ \hline
                    0.10   & -3.3990    & -4.1895    \\ \hline
                    0.12   & -3.5348    & -4.2434    \\ \hline
                    0.14   & -3.6705    & -4.2973    \\ \hline
                    0.16   & -3.8063    & -4.3512    \\ \hline
                    0.18   & -3.9421    & -4.4050    \\ \hline
                    0.20   & -4.0779    & -4.4589    \\ \hline
    \end{tabular}
    \label{table:D1}
\end{table}

\begin{figure}
    \centerline{\includegraphics[width=14cm, height=8.4cm]{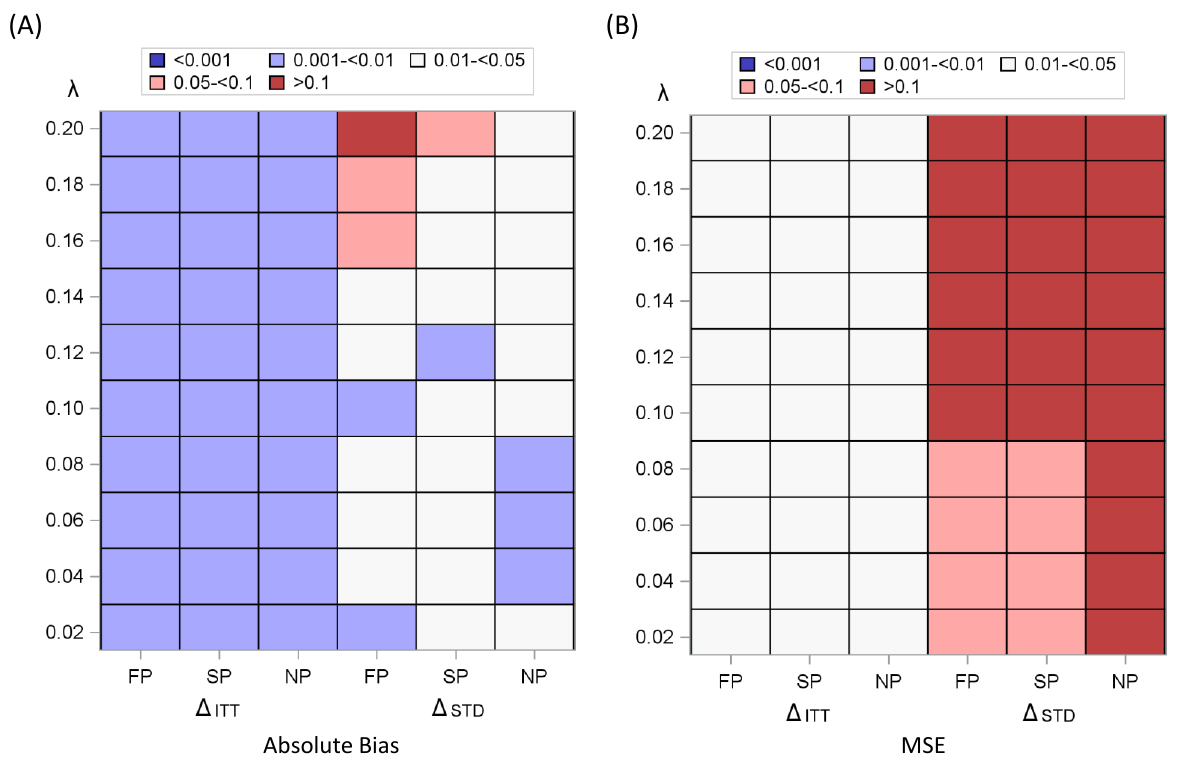}}
    \caption{\textbf{Absolute Bias and MSE for Simulated Data When $R$ and $M$ Interacts (SS = 1,000)}}
    \label{fig:figD1}
\end{figure}

\begin{figure}
    \centerline{\includegraphics[width=14cm, height=8.4cm]{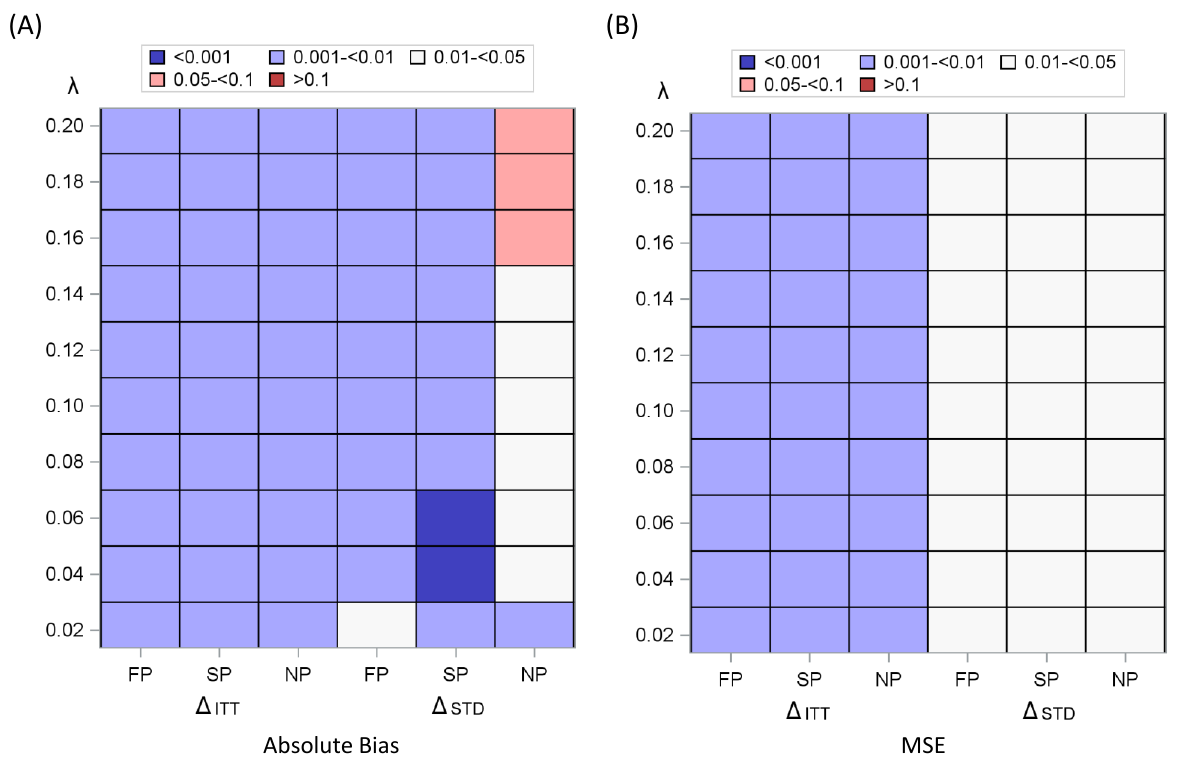}}
    \caption{\textbf{Absolute Bias and MSE for Simulated Data When $R$ and $M$ Interacts (SS = 5,000)}}
    \label{fig:figD2}
\end{figure}

\noindent\textbf{Simulation Results}\newline
This simulation studies were conducted to assess the performance of estimators for $\Delta_{ITT}$ and $\Delta_{STD}$ under different intensities of the interaction effect ($\lambda$), separately for the two sample sizes ($SS=1,000$ and $SS=5,000$), and using 1,000 replications. We compared full-parametric, semi-parametric, and non-parametric estimation methods. Performance was measured by absolute bias and MSE, with results visualized in Figures \ref{fig:figD1} and \ref{fig:figD2} and detailed numerically in Table \ref{table:D2}.

First, for the intention-to-treat effect ($\Delta_{ITT}$), the absolute bias and MSE remain stable across all estimation methods and values of $\lambda$ for a given sample size. For instance, absolute bias consistently falls within a range of 0.001 to 0.01 for both sample sizes (N=1,000 and N=5,000). The MSE values, which decrease with increased sample size, range from 0.01 to 0.05 for SS=1,000 and from 0.001 to 0.01 for SS=5,000. This pattern aligns with the expectation that estimates of $\Delta_{ITT}$ become less variable as the sample size increases.

Second, the performance for the standardized treatment effect ($\Delta_{STD}$) is slightly moderated by the complexity of the CATE function. In general, the absolute bias for $\Delta_{STD}$ increases with the intensity of the $R \times M$ interaction effect. At a smaller sample size (SS=1,000), the non-parametric method demonstrates a lower absolute bias than the parametric and semi-parametric approaches, likely due to its flexibility in capturing the complex CATE function. However, this comes at the cost of a higher MSE. At the larger sample size (SS=5,000), the full-parametric and semi-parametric methods, which correctly specify the CATE function, perform similarly and achieve a lower absolute bias than the non-parametric method. The MSE for $\Delta_{STD}$ is comparable across all three methods, ranging from 0.01 to 0.05.

In summary, under this revised simulation setting, the proposed generic identification method performs robustly across varying intensities of the second-order effect modification. It yields levels of absolute bias and MSE for both $\Delta_{ITT}$ and $\Delta_{STD}$ that are comparable to the model without an interaction term in the CATE function (see Table \ref{table:C2}). As anticipated, the sample size is critical for the accurate and precise estimation of $\Delta_{STD}$, particularly when the CATE function is complex. This mirrors a common principle in clinical research: identifying effect modifiers (i.e., interaction effects) typically requires a large sample to achieve sufficient statistical power.

\begin{table}[!ht]
    \renewcommand{\thetable}{D2}
    \centering
    \caption{\textbf{Simulation Results for $\Delta_{ITT}$ and $\Delta_{STD}$ In the Revised Scenario 1 When $R$ and $M$ Interacts}}
    \renewcommand{\arraystretch}{1.2}
    \resizebox{\textwidth}{!}{
    \begin{tabular}{|c|c|c|c|c|c|c|c|c|c|c|c|c|}
    \hline
        \textbf{} & \multicolumn{4}{c|}{\textbf{Full Parametric}} & \multicolumn{4}{c|}{\textbf{Semi-Parametric}} & \multicolumn{4}{c|}{\textbf{Non-Parametric}} \\ \hline
        \textbf{} & \multicolumn{2}{c|}{$\Delta_{ITT}$} & \multicolumn{2}{c|}{$\Delta_{STD}$} & \multicolumn{2}{c|}{$\Delta_{ITT}$} & \multicolumn{2}{c|}{$\Delta_{STD}$} & \multicolumn{2}{c|}{$\Delta_{ITT}$} & \multicolumn{2}{c|}{$\Delta_{STD}$} \\ \hline
        \textbf{$\lambda$} & Abs Bias & MSE & Abs Bias & MSE & Abs Bias & MSE & Abs Bias & MSE & Abs Bias & MSE & Abs Bias & MSE \\ \hline
        \multicolumn{13}{|c|}{\textbf{SS = 1,000}} \\ \hline
        0.02 & 0.0047 & 0.0175 & 0.0021 & 0.0737 & 0.0041 & 0.0177 & 0.0285 & 0.0756 & 0.0054 & 0.0176 & 0.0119 & 0.1563 \\ \hline
        0.04 & 0.0046 & 0.0173 & 0.0144 & 0.0772 & 0.0040 & 0.0176 & 0.0410 & 0.0792 & 0.0051 & 0.0175 & 0.0057 & 0.1588 \\ \hline
        0.06 & 0.0045 & 0.0173 & 0.0224 & 0.0833 & 0.0041 & 0.0175 & 0.0499 & 0.0848 & 0.0046 & 0.0174 & 0.0013 & 0.1615 \\ \hline
        0.08 & 0.0044 & 0.0172 & 0.0155 & 0.0901 & 0.0039 & 0.0175 & 0.0456 & 0.0929 & 0.0045 & 0.0174 & 0.0065 & 0.1647 \\ \hline
        0.10 & 0.0040 & 0.0172 & 0.0010 & 0.1024 & 0.0035 & 0.0174 & 0.0272 & 0.1036 & 0.0042 & 0.0174 & 0.0129 & 0.1680 \\ \hline
        0.12 & 0.0036 & 0.0172 & 0.0246 & 0.1182 & 0.0031 & 0.0174 & 0.0093 & 0.1168 & 0.0040 & 0.0174 & 0.0193 & 0.1708 \\ \hline
        0.14 & 0.0035 & 0.0172 & 0.0437 & 0.1326 & 0.0030 & 0.0175 & 0.0100 & 0.1293 & 0.0038 & 0.0174 & 0.0262 & 0.1744 \\ \hline
        0.16 & 0.0033 & 0.0173 & 0.0697 & 0.1441 & 0.0028 & 0.0175 & 0.0316 & 0.1409 & 0.0035 & 0.0175 & 0.0331 & 0.1787 \\ \hline
        0.18 & 0.0031 & 0.0174 & 0.0878 & 0.1423 & 0.0025 & 0.0176 & 0.0499 & 0.1424 & 0.0032 & 0.0176 & 0.0403 & 0.1831 \\ \hline
        0.20 & 0.0028 & 0.0175 & 0.1073 & 0.1360 & 0.0023 & 0.0178 & 0.0743 & 0.1367 & 0.0029 & 0.0177 & 0.0473 & 0.1880 \\ \hline
        \multicolumn{13}{|c|}{\textbf{SS = 5,000}} \\ \hline
        0.02 & 0.0040 & 0.0033 & 0.0150 & 0.0135 & 0.0040 & 0.0033 & 0.0091 & 0.0134 & 0.0038 & 0.0033 & 0.0074 & 0.0298 \\ \hline
        0.04 & 0.0039 & 0.0033 & 0.0058 & 0.0130 & 0.0040 & 0.0033 & 0.0001 & 0.0130 & 0.0038 & 0.0033 & 0.0128 & 0.0304 \\ \hline
        0.06 & 0.0040 & 0.0033 & 0.0061 & 0.0136 & 0.0040 & 0.0032 & 0.0001 & 0.0136 & 0.0039 & 0.0032 & 0.0182 & 0.0312 \\ \hline
        0.08 & 0.0040 & 0.0033 & 0.0091 & 0.0137 & 0.0040 & 0.0032 & 0.0030 & 0.0137 & 0.0039 & 0.0032 & 0.0241 & 0.0320 \\ \hline
        0.10 & 0.0040 & 0.0033 & 0.0093 & 0.0139 & 0.0041 & 0.0032 & 0.0030 & 0.0138 & 0.0039 & 0.0032 & 0.0301 & 0.0330 \\ \hline
        0.12 & 0.0041 & 0.0033 & 0.0093 & 0.0141 & 0.0041 & 0.0032 & 0.0030 & 0.0140 & 0.0039 & 0.0032 & 0.0364 & 0.0342 \\ \hline
        0.14 & 0.0040 & 0.0033 & 0.0093 & 0.0143 & 0.0040 & 0.0032 & 0.0029 & 0.0142 & 0.0038 & 0.0032 & 0.0431 & 0.0355 \\ \hline
        0.16 & 0.0040 & 0.0033 & 0.0093 & 0.0145 & 0.0041 & 0.0032 & 0.0028 & 0.0144 & 0.0038 & 0.0032 & 0.0503 & 0.0368 \\ \hline
        0.18 & 0.0041 & 0.0033 & 0.0093 & 0.0147 & 0.0041 & 0.0033 & 0.0026 & 0.0146 & 0.0038 & 0.0033 & 0.0574 & 0.0384 \\ \hline
        0.20 & 0.0041 & 0.0033 & 0.0093 & 0.0149 & 0.0041 & 0.0033 & 0.0025 & 0.0148 & 0.0038 & 0.0033 & 0.0646 & 0.0402 \\ \hline
    \end{tabular}}
    \label{table:D2}
\end{table}

\subsection{Practical Justification for a First-Order Effect Modifiers CATE Function}
\label{sec:SectionD3}
Although the proposed method has been theoretically and empirically validated to handle higher-order effect modifiers, their presence in a RCT is unlikely when a very strong primary modifier, such as placebo response ($R$), is already known. The improbability of a powerful higher-order modifier ($M$) interacting with $R$ (e.g., an $R \times M$ interaction) arises from two primary considerations: the diminished of residual treatment effect heterogeneity and the conceptual hierarchy of effects.

The argument from variance partitioning provides a key statistical rationale. In the CATE function, the total heterogeneity in treatment effect is finite. A "very strong" effect modifier, by definition, accounts for a substantial portion of this variance. The variable $R$ (placebo response) is identified as such because it explains a dominant share of the variation in treatment response across the study population. For another variable $M$ to act as a strong higher-order modifier, it would need to explain a significant fraction of the residual treatment effect heterogeneity within the subgroups defined by $R$. After accounting for the dominant effect of $R$, the residual variance is necessarily small. Consequently, the potential for any other variable to account for a large, independent portion of this diminished residual variance is limited. In essence, a very strong primary effect modifier "consumes" much of the explainable heterogeneity, thereby precluding the existence of another variable with a similarly strong, interactive influence.

From a clinical and biological perspective, the identification of a very strong placebo response ($R$) points to a primary, dominant pathway rooted in neuro-biological mechanisms of patient expectation and perception. For a second variable $M$ to create a powerful interaction, meaning it drastically alters the efficacy of the active treatment specifically within different levels of placebo response, it would require a biological pathway that is not only potent but also largely independent of this primary neuro-biological pathway. Such a scenario is biologically less probable. It is more conceptually believable that the strongest modifier represents the principal driver of heterogeneity, and that other influential factors operate with a lesser, likely additive effect rather than a powerful multiplicative one.

In summary, while the proposed method is robust to the potential for higher-order interactions, a scenario featuring both a very strong primary effect modifier (placebo response $R$) and a powerful higher-order modifier ($M$) is highly unlikely. This conclusion is grounded in the statistical constraint of limited residual heterogeneity and the biological implausibility of two independent, dominant pathways operating in parallel. Therefore, in practical analysis, it is methodologically sound to prioritize the comprehensive characterization of the primary, first-order strong effect modifier rather than searching for complex higher-order interactions that are unlikely to be substantial.

\end{document}